\documentclass[a4paper, 12pt]{article}
\usepackage[margin=1in]{geometry}
\usepackage{amsmath}
\usepackage{bm}
\usepackage{tikz}
\usetikzlibrary{arrows,automata}
\usepackage{enumerate}
\usepackage{graphics}
\usepackage{epstopdf}
\usepackage{epsfig}
\usepackage{amssymb}
\usepackage{dsfont}
\usepackage{microtype}
\usepackage[colorlinks,citecolor=blue,urlcolor=red,bookmarks=false,hypertexnames=true]{hyperref} 
\usetikzlibrary{snakes}
\usepackage[normalem]{ulem}
\usepackage{caption}
\usepackage{subcaption}
\usepackage{caption}
\usepackage{breakcites}
\usepackage{xcolor}
\usepackage{booktabs}
\newcommand{\ra}[1]{\renewcommand{\arraystretch}{#1}}

\providecommand{\keywords}[1]
{
	\small
	\textbf{\textit{Keywords:}} #1
}

\usepackage{authblk}
\usepackage{natbib}

\title{Bayesian Tensor Factorisations for Time Series of Counts}
\author[1]{Zhongzhen Wang}
\author[1,2]{Petros Dellaportas\thanks{Corresponding author. Email: p.dellaportas@ucl.ac.uk}}
\author[3]{Ioannis Kosmidis}
\affil[1]{University College London, London, UK}
\affil[2]{Athens University of Economics and Business, Athens, Greece}
\affil[3]{University of Warwick, Coventry, UK}
\usepackage[framemethod=tikz]{mdframed}

\begin{document}
	\title{Bayesian Tensor Factorisations for Time Series of Counts}
	
	\maketitle

	\begin{abstract}
We propose a  flexible nonparametric Bayesian modelling framework for multivariate time series of count data based on tensor factorisations. Our models can be viewed as infinite 
state space Markov chains of known maximal order with non-linear serial dependence through the introduction of appropriate latent variables. Alternatively, our models can be viewed as Bayesian
hierarchical models with conditionally independent Poisson distributed observations.  Inference about the important lags and their complex interactions is achieved via MCMC.  When the observed counts are large, we deal with the resulting computational complexity of Bayesian inference via a two-step inferential strategy based on an initial analysis of a training set of the data.  Our methodology is illustrated using simulation experiments and analysis of real-world data. 
	\\
	\end{abstract}

\section{Introduction}
We consider a time-index sequence of multivariate random variables of size $T$, $\{y_t\}_{t=1}^T$, taking values in  $\{0,1,\ldots\}$. We build a non-parametric model by (i) assuming that the transition probability law of the sequence $\{ y_{t} \}$ 
conditional on the filtration up to time $t-1$, $\mathcal{F}_{t-1}$, is that of a Markov chain of maximal order $q$,
(ii) allowing non-linear dependence of the values at the previous $q$ time points and (iii) incorporating complex interactions between lags. 

We propose a Bayesian model for multivariate time series of counts based on tensor factorisations. Our development is inspired by \cite{yang2016bayesian} and \cite{sarkar2016bayesian}.  \cite{yang2016bayesian}  introduced conditional tensor factorisation models that lead to parsimonious representations of transition probability vectors together with a simple, powerful Bayesian hierarchical formulation based on latent allocation variables.  This framework has been exploited in \cite{sarkar2016bayesian}  to build a nonparametric Bayesian model for categorical data together with an efficient MCMC inferential framework.  We adopt the ideas and methods of these papers to build flexible models for time series of counts.  The major difference that distinguishes our work to \cite{sarkar2016bayesian} is that, unlike categorical data, we deal with time series that are  infinite, rather than finite, state space Markov chains.  The resulting computational complexity of our proposed model is grown as the observed counts become larger, so we propose a two-step inferential strategy in which an initial, training part of the time series data, is utilized to facilitate the inference and prediction of the rest of the data. 

A common way to analyse univariate time series of counts is by assuming that the conditional probability distribution of $y_t \mid y_{t-1}, \dots, y_{t-q}$ can be expressed as a Poisson density with rate 
$\lambda_{t}$ that depends either on previous counts $y_{t-1}, \dots, y_{t-q}$ or previous intensities $\lambda_{t-1}, \dots, \lambda_{t-q}$. For example, 
one such popular model is the Poisson autoregressive model (without covariates) of order $q$, PAR($q$):
\begin{equation}
\begin{aligned}
y_t  &\sim \textrm{Poisson}(\lambda_t), \\
\log (\lambda_t) &= \beta_0 + \sum_{i= 1}^{q} \beta_i \log (y_{t-i} + 1) 
\end{aligned}
\label{PAR}
\end{equation}
where $\beta_0, \beta_1, \dots, \beta_q$ are unknown parameters; see  \cite{cameron2001essentials}.  \cite{grunwald1995unified}, \cite{grunwald1997some} and \cite{ fokianos2011some} discuss the modelling and properties of a PAR($1$) process. \cite{brandt2001linear} generalise PAR(1) to a PAR($q$) process and apply it to the modelling of presidential vetoes in the United States. \cite{kuhn1994use} adopt such processes to model the counts of child injury in Washington Heights. When we deal with $M$ distinct time series of counts, the PAR($q$) model is written, for $m=1,\ldots,M$, as 
\begin{equation}
\begin{aligned}
y_{m,t}  &\sim \textrm{Poisson}(\lambda_{m,t}), \\
\log (\lambda_{m,t}) &= \beta_0 + \sum_{m=1}^{M} \sum_{i= 1}^{q} \beta_{i,m} \log (y_{m,t-i} + 1);
\end{aligned}
\label{MPAR}
\end{equation}
%where the above formulation is written for each of the  $m=1,2,\ldots,M$ univariate time series, 
see, for example, \cite{liboschik2015tscount}. In the above equation, $q$ is fixed for each $m=1,\ldots,M$. We will use
this model formulation as a benchmark for comparison against our proposed methodology.
Other approaches to modelling time series of counts include the integer-valued generalised autoregression conditional heteroscedastic models  \cite{heinen2003modelling, weiss2014ingarch} and the integer-valued autoregression processes \cite{al1987first}.  We have not dealt with these models here because a proper Bayesian evaluation of their predictive performance requires a challenging Bayesian inference task which is beyond the scope of our work.

The rest of the paper is organised as follows. We specify our model in Section 2, followed by estimation and inference details in Section 3. Simulation experiments and applications are provided in Section 4 and 5, respectively. 

\section{Model Specification}

\subsection{The Bayesian tensor factorisation model}
\subsubsection{Univariate time series}
\label{univariate}
We build a probabilistic model by assuming that the transition probability law of $y_{t}$ conditional on $\mathcal{F}_{t-1}$ is that of a Markov chain of maximal order $q$:
\begin{equation}
\begin{aligned}
p(y_{t} &\mid \mathcal{F}_{t-1}) =p(y_{t} \mid \{y_{t-j}\}_{j \in {[1,q]}}), 
\end{aligned}
\label{MC_q_US}
\end{equation}
for $t \in {[q+1, T]}$ where the set containing all integers from $i$ to $j$ is denoted as ${[i,j]}$. This formulation includes the possibility that only a subset of the previous $q$ values affects $y_t$. We follow \cite{sarkar2016bayesian} and introduce a series of latent variables as follows. First, let $k_j$ denote the maximal number of clusters that the values of $y_{t-j}$ can be separated into for predicting $y_t$. 
To demonstrate the use of $k_j$ we present a simple example. Assume that $y_t$ depends only on $y_{t-1}$  and the relationship in which the observed values of $y_{t-1}$ affect the density of $y_t$ is based on the following stochastic rule: if $y_{t-1} > 1$ then $y_t \sim \textrm{Poisson}(1)$  and if $y_{t-1} \le 1$ then $y_t\sim \textrm{Poisson}(2)$.  Then $k_1 = 2$ since the values of $y_{t-1}$ are separated into two clusters that determine the distribution of $y_t$. Note that if  $k_j = 1$ the value of $y_{t-j}$ does not affect the density of $y_t$. The collection of all these latent variables 
$K:= \{k_{j}\}_{j \in {[1, q]}}$ determines how past values of the time series affect the distribution of $y_t$.   

We also define a collection of time-dependent latent allocation random variables $Z_t:= \{z_{j,t}\}_{j\in {[1,q]}}$ where $z_{j,t}$ specifies which of the $k_j$ clusters of $y_{t-j}$
affects $y_t$. We will write $Z_t=H$ meaning that all latent variables in $Z_t$ equal to another collection of latent variables 
$H:= \{h_{j}\}_{j\in {[1,q]}}$ that do not depend on $t$.
Finally, denote the collection $\mathcal{H} := \{h_{j} \in {[1, k_{j}]}, j \in {[1,q]}\}$ that depends on $K$.
The connection among $Z_t$, $H$ and $\mathcal{H}$ is that for any $t \in {[q+1, T]}$, $Z_t$ is sampled with value $H \in \mathcal{H}$. 

We are now in a position to define our model. Let $\lambda_{Z_t}$ be the Poisson rate for generating $y_t$ given the random variable $Z_t$.  The conditional transition probability law (\ref{MC_q_US}) can be written as a Bayesian hierarchical model, for $j\in {[1,q]}$, $H \in \mathcal{H}$ and $t \in {[q+1, T]}$, as
\begin{align}
&y_{t} \mid Z_t = H \sim \textrm{Poisson}(\lambda_H), \label{US_Bayesian tensor factorisation_lv1}\\
&z_{j,t} \mid y_{t-j}  \sim \textrm{Multinomial}\left({[1,k_{j}]}, \big( \pi_{1}^{(j)}(y_{t-j}), \dots, \pi_{k_{j}}^{(j)}(y_{t-j}) \big) \right). \label{US_Bayesian tensor factorisation_lv2}
\end{align}
Expressions (\ref{US_Bayesian tensor factorisation_lv1}) and (\ref{US_Bayesian tensor factorisation_lv2}) imply that
\begin{equation}
\begin{aligned}
& p(y_{t} \mid \{y_{t-j}\}_{j \in {[1,q]}}) = \sum_{H \in \mathcal{H}} \textrm{PD}(y_{t}; \lambda_{H})\prod_{j \in {[1,q]}} \pi_{h_{j}}^{(j)}(y_{t-j}).
\end{aligned}
\label{US_Bayesian tensor factorisation}
\end{equation} 
with constraints
$\lambda_{H} \ge 0$ for any $ H \in \mathcal{H}$ and 
$\sum_{h_{j} = 1}^{k_{j}}\pi^{(j)}_{h_{j}}(y_{t-j}) = 1$
for each combination of $(j, y_{t-j})$.
Multinomial(${[1,k]}, \pi$) is a multinomial distribution selecting a value from ${[1,k]}$ with a probability vector $\pi$. 
The formulation (\ref{US_Bayesian tensor factorisation}) is referred to as a conditional tensor factorisation with the Poisson density PD($y_t; \lambda_{H}$) being the core tensor;  see \cite{harshman1970foundations, harshman1994parafac, tucker1966some, de2000multilinear} for a description of tensor factorisations. It can also be interpreted as a Poisson mixture model with $\prod_{j \in {[1,q]}} \pi_{h_{j}}^{(j)}(y_{t-j})$ being the  mixture weights that depend on previous values of $y_t$. 
%This mixture model has $\prod_{j\in {[1,q]}} k_j$ parameters that need to be estimated.\textcolor{red}{(delete)}

A more parsimonious representation for our tensor factorisation model is obtained by adopting a  Dirichlet process for Poisson rates $\lambda_{H}$. Independently, for each $H \in \mathcal{H}$, we use the stick-breaking construction introduced by \cite{sethuraman1994constructive} in which
\begin{equation}
\lambda_{H} \sim \sum_{l=1}^{\infty} \pi_l^{\ast} \delta(\lambda_l^{\ast}),
\label{ind_lambda_US}
\end{equation}
where $\delta(.)$ is a Dirac delta function and independently, for $l \in {[1,\infty)}$,
\begin{equation*}
\pi^{\ast}_{l} = V_l \prod_{s=1}^{l-1}(1-V_{s}), \enspace V_l \sim \textrm{Beta}(1, \alpha_0), \enspace \lambda_l^{\ast}  \sim \textrm{Gamma}(a, b) 
\end{equation*}
where $\lambda_l^{\ast}$  represents a label-clustered Poisson rate. By letting $\mathcal{Z}_{Z_t}^{\ast}$ denote the label of the cluster that $Z_t$ belongs to at time $t \in {[q+1, T]}$,  we complete the model formulation as
\begin{equation}
\begin{aligned}
&p(\mathcal{Z}^{\ast}_{H} = l) = \pi^{\ast}_{l}, \enspace \textrm{independently for each } H \in \mathcal{H} , \\
&(\lambda_{H} \mid \mathcal{Z}^{\ast}_{H} = l) = \lambda_l^{\ast}, \\
&(\mathcal{Z}^{\ast}_{Z_t} \mid Z_t = H ) = \mathcal{Z}^{\ast}_{H}, \\
&(y_{t} \mid \mathcal{Z}^{\ast}_{Z_t} = l) \sim \textrm{Poisson}(\lambda^{\ast}_l).
\end{aligned}
\label{label_US}
\end{equation}

\subsubsection{Multivariate time series}
The model of the previous section can be readily extended to deal with multivariate responses $\{ Y_t\}_{t=1}^T$, where $Y_t = ( y_{1,t}, \dots, y_{M,t} )^\top$ taking values in $\mathbb{N}_0$.   The idea is similar to the way the univariate PAR model (\ref{PAR}) is generalised to its multivariate counterpart (\ref{MPAR}). We assume that the transition probability  for any $\tau \in {[1, M]}$ and $t \in {[q+1, T]}$ is
\begin{equation}
\begin{aligned}
p(y_{\tau,t} &\mid \mathcal{F}_{t-1}) =p(y_{\tau,t} \mid \{y_{m,t-j}\}_{m \in {[1, M]}, j\in {[1,q]}}).
\end{aligned}
\label{MC_q_MS}
\end{equation} 
The idea is that each univariate time series may depend on all or some of the $q$ previous values of all, or some, univariate time series. Model (\ref{MC_q_MS}) assumes
that conditional on past $q$ values of all time series before time $t$, the $M$ univariate random variables at time $t$ are independent.  The formulation requires $M$ different latent variables for each dimension 
but, other than that, its specific details have no essential difference from those in the univariate case.

\subsubsection{Two-step inference for large counts}

Imagine that based on observed data $\{y_t\}_{t=1}^T$, one has to recursively forecast future observations $y_{T+1}, y_{T+2}, \ldots$.
Clearly, the observed values of $\{y_t\}_{t=1}^T$ determine the form of our models in Sections 2.1.1 and 2.1.2 and as a result of this construction we may face the unfortunate situation in which a count that has been unobserved up to time $T$ appears in the future observations. This problem can be solved by re-estimating the model but in cases where this is not desirable, we propose the following solution. We separate $\{y_t\}_{t=1}^T$ into two segments of size $T_1$ and $T_2$, representing the size of \textit{pre-training} dataset and \textit{training} dataset, respectively, so  $\{y_t\}_{t \in {[1, T_1]}}$ and $\{y_t\}_{t\in {[T_1+1, T_1+T_2]}}$ are the corresponding observations in these sets. We aim to use the pre-training dataset to cluster all the counts in time series and  the training dataset to model the time series with labelled counts.

We first define a collection of latent variables $\{w_{1:c-1}, \mu_{1:c}, c\}$
that models the pre-training data $\{y_t\}_{t \in {[1, T_1]}}$ as
\begin{equation}
\begin{aligned}
p(y_{t}\mid w_{1:c-1}, \mu_{1:c}, c) = \sum_{i=1}^{c} w_{i} \textrm{PD}(y_{t}; \mu_{i})
\end{aligned}
\label{uni_Poi_mix}
\end{equation}
for any $t \in {[1, T_1]}$, $0 < w_{i} < 1$, $\sum_{i=1}^{c}w_{i} = 1$, $\mu_{i} \ge 0$.  Thus, (\ref{uni_Poi_mix}) assumes that any $y_{t}$ in the pre-training dataset is distributed as a finite mixture of Poisson distributions with $c$ components, weights $w_{i}$ and intensities $\mu_{i}$.  The usual latent structure for such mixture models assumes indicator variables $d_{t}$ representing the estimated label of the mixture component that $y_{t}$ belongs to, so $p(d_{t} = i) = w_i$  for all $i \in {[1,c]}$. 

We exploit this finite mixture clustering of the pre-training dataset to build our model for the training dataset. 
We define another collection of latent variables as $D_t = \{ d_{j,t} \}_{j\in {[1, q]}}$ and
by setting $d_{j,t} = d_{t-j}$ for all $j\in {[1, q]}$ and $t\in {[T_1+1+q, T_1+T_2]}$.
We then build a probabilistic model for the training dataset by assuming that the transition probability law of the sequence $\{ y_{t} \}_{t \in {[T_1+1+q, T_1+T_2]}}$ conditional on $\mathcal{F}_{t-1}$ is that of a probabilistic model of this target sequence conditional on $D_t$ . That is,  we have
\begin{equation}
\begin{aligned}
p(y_{t} &\mid \mathcal{F}_{1:t-1}) =p(y_{t} \mid D_t).
\end{aligned}
\label{MC_q_1}
\end{equation} 

The conditional transition probability law (\ref{MC_q_1}) can then be written as a Bayesian hierarchical model, for $j \in {[1, q]}$ and $t \in {[T_1+1+q, T_1+T_2]}$, as
\begin{align}
&y_{t} \mid Z_t = H \sim \textrm{Poisson}(\lambda_{H}), \label{UG_Bayesian tensor factorisation_lv1}\\
&z_{j,t} \mid d_{j,t}  \sim \textrm{Multinomial}\left({[1,k_{j}]}, \big\{ \pi_{1}^{(j)}(d_{j,t}), \dots, \pi_{k_{j}}^{(j)}(d_{j,t}) \big\}\right). \label{UG_Bayesian tensor factorisation_lv2}
\end{align}
(\ref{UG_Bayesian tensor factorisation_lv1}) and (\ref{UG_Bayesian tensor factorisation_lv2}) immediately imply that
\begin{equation}
\begin{aligned}
& p(y_{t} \mid D_t) = \sum_{H \in \mathcal{H}} \textrm{PD}(y_{t}; \lambda_{H}) \prod_{j \in {[1,q]}} \pi_{h_{j}}^{(j)}(d_{j,t}).
\end{aligned}
\label{UG_Bayesian tensor factorisation}
\end{equation} 
with constraints
$\lambda_{H} \ge 0$ for any $H \in \mathcal{H}$ and 
$ \sum_{h_{j} = 1}^{k_{j}}\pi^{(j)}_{h_{j}}(d_{j,t}) = 1$
for each combination of $(j, d_{j,t})$.
It is clear that (\ref{UG_Bayesian tensor factorisation}) is equivalent to (\ref{UG_Bayesian tensor factorisation_lv1}) and (\ref{UG_Bayesian tensor factorisation_lv2}). From (\ref{UG_Bayesian tensor factorisation}) the expectation of $y_{t}$ conditional on $D_t$ is
\begin{equation}
\begin{aligned}
\mathbb{E}(y_{t} \mid D_t) = \sum_{H \in \mathcal{H}} \lambda_{H} \prod_{j \in {[1, q]}} \pi_{h_{j}}^{(j)}(d_{j,t}).
\end{aligned}
\label{predictive_mean}
\end{equation}
The rest of the model which utilises the stick-breaking process for $\lambda_H$ is similar to the one used in Section \ref{univariate}.

\subsubsection{Priors}
We assign independent priors on $\pi^{(j)}(d_{j,t})$ as
\begin{equation*}
\pi^{(j)}(d_{j,t}) = \{ \pi^{(j)}_1(d_{j,t}), \dots, \pi^{(j)}_{k_{j}}(d_{j,t}) \} \sim \textrm{Dirichlet}(\gamma_{j}, \dots, \gamma_{j}),
\label{pi_UG}
\end{equation*}
with $\gamma_{j} = 0.1$. Also, we follow  \cite{sarkar2016bayesian} and set priors
\begin{equation*}
p(k_{j} = \kappa)  \propto \exp(-\varphi j\kappa), 
\label{k_j_UG}
\end{equation*}
where $j \in {[1, q]}$, $\kappa \in {[1,c]}$.
Notice that $\varphi$ controls $p(k_{j} = \kappa)$ and the number of important lags for the proposed conditional tensor factorisation; for all our experiments throughout this paper, we set $\varphi = 0.5$.  Following  \cite{viallefont2002bayesian}, we place for the Gamma density of $\lambda_l^{\ast}$ parameters $a$ as the mid-range of $y_{t}$ in the training dataset
	$a = \frac{1}{2} [\max (\{y_t\}_{t \in {[T_1+1+q, T_1+T_2]} }) - \min (\{y_t\}_{t \in {[T_1+1+q, T_1+T_2]} })]$ and $b = 1$.  We set $\alpha_0 = 1$ for the Beta prior to $V_l$. Finally, we truncate the series (\ref{ind_lambda_US}), by assuming 
	\begin{equation*}
	\lambda_{H} \sim \sum_{l=1}^{L} \pi_l^{\ast} \delta(\lambda_l^{\ast}),
	\label{ind_lambda_UG}
	\end{equation*}
	and we set  $L=100$.

\section{Estimation and Inference}

The joint density of the general model of Section {\color{blue} 2.2.3} can be expressed as \newline
$p(y, Z, \mathcal{Z}^{\ast}, D, \lambda^{\ast}, \pi^{\ast}, \pi_K)$, where $D = \{D_t\}_{t\in {[T_1+1, T_1+T_2]}}$ and $K = \{k_j\}_{j \in {[1,q]}}$. The Poisson mixture model in the pre-training set is estimated with the MCMC algorithm of  \cite{marin2005bayesian}.  For any $t>T_1$ we then estimate 
$d_{t} = \arg_i \max \textrm{PD}(y_{t}, \mu_i)$, $i \in {[1,c]}$. Our BTF model has a finite number of mixture components with an unknown number of components due to the randomness of the random variable matrix $K$.  We follow \cite{yang2016bayesian} 
and estimate  $K$ separately through a stochastic search variable selection \cite{george1997approaches} based on approximated marginal likelihood. As \cite{yang2016bayesian}  point out, such an approach is helpful since it fixes the numbers of inclusions of the tensor and the sampling process  of $K$ can indicate whether a predictor is important. The rest of the inference proceeds by sampling all other random variables conditional on $K$ and $D$ through MCMC.  

\subsection{MCMC for finite Poisson mixtures} 
We follow the procedure in \cite{marin2005bayesian}. $\{y_{t}\}_{t\in \mathbb{Z}_{[1, T_1]}}$ is a mixture of $c$ univariate Poisson distributions with density $\sum_{i=1}^{c} w_i \textrm{PD}(y_t;\mu_i)$, $\{w_i\}_{i \in \mathbb{Z}_{[1,c]}}$ are weights with $\sum_{i=1}^c w_i = 1$ and $\{\mu_i\}_{i \in \mathbb{Z}_{[1,c]}}$ are the corresponding Poisson rates. By setting the priors as
$\mu_i \sim \textrm{Gamma}(1, 1)$, $\enspace \{w_i\}_{i \in \mathbb{Z}_{[1,c]}} \sim \textrm{Dirichlet}(1, \dots, 1),$
the corresponding Gibbs sampler is as follows: (i)
Generate the label of $y_t$, $\iota_{t}$, for $t \in \mathbb{Z}_{[1, T_1]}$, $i \in \mathbb{Z}_{[1,c]}$ as 
$		p(\iota_{t} = i) \propto w_i \left( \mu_i \right)^{y_{t}} \exp \left( -\mu_i \right)$ and set $n_i = \sum_{t \in \mathbb{Z}_{[1, T_1]}} \mathds{1}_{\iota_{t} = i}$ and $\mathcal{I}_i = \sum_{t \in \mathbb{Z}_{[1, T_1]}}\mathds{1}_{\iota_{t} = i} y_{t}$
		(ii) Generate 
		$\{w_i\}_{i \in \mathbb{Z}_{[1,c]}} \sim  \textrm{Dirichlet}(1 + n_1, \dots, 1 + n_{c})$
		and (iii) for $i \in \mathbb{Z}_{[1,c]}$, generate 
		$\mu_i \sim \textrm{Gamma}(1 + \mathcal{I}_i, 1 + n_i).$  
	
\subsection{Important lags selection} 
Important lags are inferred by the variable $K = \{k_{j}\}_{ j\in \mathbb{Z}_{[1, q]}}$. The basic
calculations are as follows.
Following \cite{sarkar2016bayesian}, the posterior of $K = \{k_{j}\}_{ j\in \mathbb{Z}_{[1, q]}}$ can be sampled as 
\begin{equation*}
	p(k_{j} | \dots) \propto \exp (-\varphi j k_{j}) \prod_{\omega=1}^{c}\frac{\Gamma(k_{j} \gamma_{j})}{\Gamma(k_{j} \gamma_{j} + n_{j,\omega})}
\end{equation*}
with 
$
    k_{j} = \max \left(\{z_{j,t}\}_{t\in \mathbb{Z}_{[T_1+1+q, T_1+T_2]}}\right), \dots, c
$
and 
$
    n_{j,\omega} = \sum_{t \in \mathbb{Z}_{[T_1+1+q, T_1+T_2]}} \mathds{1} \{d_{j,t} = \omega. \}
$
The levels of $d_{j,t}$ are partitioned into $k_{j}$ clusters $\{ C_{j,r} : r = 1, \dots, k_{j} \}$ with each cluster $C_{j,r}$ assumed to correspond to its own latent class $h_{j} = r$. With independent Dirichlet priors on the mixture kernels $\lambda_H \sim \textrm{Gamma}(a, b)$ marginalised out, the likelihood of our targeted response $\{y_{t}\}_{t\in\mathcal{T}_2^{\ast}}$ conditional on the cluster configuration $C = \{C_{j,r} : j \in \mathbb{Z}_{[1,q]}, r \in \mathbb{Z}_{[1, k_j]}\}$ is given by 
\begin{align*}
	&p(\{y_{t}\}_{t\in \mathbb{Z}_{[T_1+1+q, T_1+T_2]}} \mid C) = \prod_{H \in \mathcal{H}}\int_{0}^{\infty}f(\{y_{ t}\}_{t\in \mathbb{Z}_{[T_1+1+q, T_1+T_2]}}\mid \lambda_H) p(\lambda_H \mid C) d \lambda_H \\
	&= \prod_{H\in\mathcal{H}}\int_{0}^{\infty} \left(\prod_{t\in \mathbb{Z}_{[T_1+1+q, T_1+T_2]}} (y_{t}\xi )! \right)^{-1} \exp\left(-(\sum_{t\in \mathbb{Z}_{[T_1+1+q, T_1+T_2]}}\xi) \lambda_H \right) \cdot \\
	&\lambda_H^{\sum_{t\in \mathbb{Z}_{[T_1+1+q, T_1+T_2]}}y_{t} \xi} \frac{1}{(1/b)^{a} \Gamma(a)} \lambda_H^{a-1} \exp(-\lambda_H b)  d \lambda_H\\
	&= \prod_{H \in\mathcal{H}} \frac{1}{(1/b)^{a} \Gamma(a)} \left(\prod_{t\in \mathbb{Z}_{[T_1+1+q, T_1+T_2]}} (y_{t}\xi)! \right)^{-1} \Gamma\left(a + \sum_{t\in \mathbb{Z}_{[T_1+1+q, T_1+T_2]}}y_{t} \xi\right) \cdot \\ 
	& \left(\sum_{t\in \mathbb{Z}_{[T_1+1+q, T_1+T_2]}}\xi + b\right)^{-(a + \sum_{t\in \mathbb{Z}_{[T_1+1+q, T_1+T_2]}}y_{t} \xi)},
	\end{align*}
	where $\xi = \mathds{1}\{ d_{1,t} \in C_{1, h_{1}}, \dots, d_{q,t} \in C_{q, h_{q}} \} $. Then the MCMC steps for $j \in \mathbb{Z}_{[1, q]}$ are: 
	(i) If $1 \le k_{j} \le c$, we propose to either increase $k_{j}$ to $(k_{j}+1)$ or decrease $k_{j}$ to $(k_{j}-1)$.
		(ii)  If an increasing move is proposed, we randomly split a cluster of $d_{j,t}$ into two clusters. We accept this move with an acceptance rate based on the approximated marginal likelihood.
		(iii)  If a decrease move is proposed, we randomly merge two clusters of $d_{j,t}$ into a single cluster. We accept this move with an acceptance rate based on the approximated marginal likelihood.
	If $K^{\ast}$ and $C^{\ast}$ are the updated model index and cluster, $\alpha(\cdot; \cdot)$ is the Metropolis-Hastings acceptance rate, $L(\cdot)$ is the likelihood function and $q(\cdot \rightarrow \cdot)$ is the proposal function, we obtain
	\begin{align*}
	\alpha(K, C; K^{\ast}, C^{\ast}) = \frac{L( \{y_{t}\}_{t\in \mathbb{Z}_{[T_1+1+q, T_1+T_2]}}, K^{\ast}, C^{\ast}) q(K^{\ast}, C^{\ast} \rightarrow K, C)}{L( \{y_{t}\}_{t\in \mathbb{Z}_{[T_1+1+q, T_1+T_2]}}, K, C) q(K, C \rightarrow K^{\ast}, C^{\ast})}.
	\end{align*}

\subsection{Full conditional densities} 
For given $D$ and $K$, denote by  $\zeta$  a generic variable that collects the variables that are not explicitly mentioned, including $y$. Then the corresponding Gibbs sampling steps are 
	\begin{itemize}
		\item Sample $\mathcal{Z}^{\ast}_H$ for each $H \in \mathcal{H}$ from
		$p(\mathcal{Z}^{\ast}_H = l \mid \zeta) \propto \pi^{\ast}_l  (\lambda_l^{\ast})^{n_H^{\ast}} \exp \left(-n_H\lambda_l^{\ast}\right)$
		where $n_H^{\ast} = \sum_{t\in \mathbb{Z}_{[T_1+1+q, T_1+T_2]}}\mathds{1} \{ Z_t = H  \} y_{t}$ and $ n_H = \sum_{t \in \mathbb{Z}_{[T_1+1+q, T_1+T_2]}} \mathds{1} \{ Z_t = H \}.  $
		\item Sample $V_l$ for $l \in \mathbb{Z}_{[1, L]}$ from
		$ V_l \mid \zeta \sim \textrm{Beta}\left(1 + \mathcal{N}^{\ast}_l, \alpha_0 + \sum_{l' > l} \mathcal{N}^{\ast}_{l'} \right)$
		where $\mathcal{N}^{\ast}_l = \sum _{H \in \mathcal{H}} \mathds{1} \{ \mathcal{Z}^{\ast}_H = l \}$, and update $\pi^{\ast}_l$ accordingly. 
		
		\item Sample each $\lambda^{\ast}_l$ with $l \in \mathcal{L}$ from
		$ \lambda^{\ast}_l  \mid \zeta \sim \textrm{Gamma}\left( a + N_H^{\ast}(l), b + N_H(l)\right), $
		where  $N_H^{\ast}(l) = \sum _{H \in \mathcal{H}} \mathds{1} \{ \mathcal{Z}^{\ast}_H = l \} n_H^{\ast}$ and  $ N_H(l) = \sum _{H \in \mathcal{H}} \mathds{1} \{ \mathcal{Z}^{\ast}_H = l \} n_H. $ 
		\item For $j \in \mathbb{Z}_{[1,q]}$ and $\omega \in \mathbb{Z}_{[1,c]}$, sample
		\begin{equation*}
		\left\{ \pi^{(j)}_{1}(\omega), \dots,  \pi^{(j)}_{k_{j}}(\omega) \right\} | \zeta \sim \textrm{Dirichlet}\{ \gamma_{j} + n_{j,\omega}(1), \dots,  \gamma_{j} + n_{j,\omega}(k_{j})\}
		\end{equation*}
		where $n_{j,\omega}(h_{j}) = \sum_{t \in \mathbb{Z}_{[T_1+1+q, T_1+T_2]}} \mathds{1} \{ z_{j,t} = h_{j}, d_{j,t} = \omega \}$. 
		
		\item Sample $z_{j,t}$ for $j \in \mathbb{Z}_{[1,q]}$ and $t \in \mathbb{Z}_{[T_1+1+q, T_1+T_2]}$ from
		\begin{equation*}
		p(z_{j,t} = h | z_{j',t} = h_{j'}, j' \ne j,  \zeta) \propto 
		\pi_{h}^{(j)}(d_{j,t})\left(\lambda^{\ast}_{\mathcal{Z}^{\ast}_{H_{\dots/ j = h}}}\right)^{y_{t}} \exp\left(-\lambda^{\ast}_{Z^{\ast}_{H_{\dots / j = h}}}\right),
		\end{equation*}
		where $H_{\dots / j = h}$ is equal to $H$ at all position except the $j$-th position taking the value $h$.
		
	\end{itemize}

\section{Simulation Experiments}

We tested our methodology with simulated data from designed experiments against the Poisson autoregressive model (\ref{PAR}) through the log predictive score calculated  in an out-of-sample (test) dataset 
$\mathfrak{T}$ of size ${\tilde T}$.  For each model the log predictive score is estimated by 
$$  \frac{-\sum_{t \in \mathfrak{T}}\sum_{i=1}^N \log \hat{p}^{(i)}(y_t) }{{\tilde T}N}$$
where  $\hat{p}^{(i)}(y_t)$ denotes the 
one-step ahead estimated transition probability of observing $y_{t \in \mathfrak{T}}$ calculated using the parameter values at the $i$-th iteration of MCMC with total $N$ iterations.  It measures the predictive accuracy of the model by assessing the quality of the uncertainty quantification.   A model predicts better when the log predictive score is smaller; see, for example, \cite{czado2009predictive}. 
For each designed scenario, we generated $10$ datasets with $5,000$ data points and out-of-sample predictive performance for all models was tested by using either the first $4,000$ or $4,500$ data points as training datasets and calculating the log predictive scores approximated via the MCMC output at the rest $1,000$ or $500$ test data points respectively.  The resulting mean log predictive score that is reported in Tables $1$-$3$  is the average log predictive score across the $10$ generated datasets. 
The pre-training dataset for the BTF model has been chosen to be the first $3,000$ points. All MCMC runs were based on the following burn-in and posterior samples respectively: $2,000$ and $5,000$ for fitting the Poisson mixtures on the pre-training dataset; $1,000$ and  $2,000$ for selecting the important lags and their corresponding number of inclusions; and $2,000$ and $5,000$ for sampling the rest of the parameters. 
Bayesian inference for Poisson autoregressive model was obtained by 'rjags' \cite{plummer2016rjags} package based on  5,000 burn-in and 10,000 MCMC samples respectively.  We first chose the order $q$ of the model by choosing among all models with maximum order up to $q+2$ using the AIC and BIC criteria. 
We set the priors for parameters as $\beta_0 \sim N(0, 10^{-6})$ and $\beta_i \sim N(0, 10^{-4})$ for any $i \in {[1,q]}$.

Table \ref{tab:Comp_logLC} presents the results of out-of-sample comparative predictive ability based on six generated Poisson autoregressive models  based on (\ref{PAR}).  Notice that when the order $q$ is high and there are only a few true coefficients, as in cases $C,E$ and $F$, the maximal order Markov structure of the BTF model achieves a comparative, satisfactory predictive performance.  Given that the data generating process is based on Poisson autoregressive models these results are very promising.

\begin{table}[h!]\centering
	\ra{1}
	\scalebox{0.78}{
		\begin{tabular}{@{}llllllllllll@{}}\toprule
		& & \multicolumn{2}{c}{Bayesian Poisson autoregression} & \multicolumn{1}{c}{BTF} \\
		\cmidrule{3-4}
		Scenarios & Data Sizes & \multicolumn{1}{c}{AIC} & \multicolumn{1}{c}{BIC} \\ \midrule
		$(A):$ $\beta_0 = 1, \beta_1 = 0.5$ & $4000:1000$ &$\bm{2.436(0.024)}$ & $\bm{2.436(0.024)}$ & $2.443(0.022)$ \\
		 & $4500:500$ &$\bm{2.441(0.031)}$ & $\bm{2.441(0.031)}$ & $2.450(0.030)$ \\
		\hline
		$(B):$ $\beta_0 = 1, \beta_7 = 0.5$ & $4000:1000$ &$2.450(0.019)$ & $\bm{2.449(0.019)}$ & $2.458(0.022)$ \\
		 & $4500:500$ &$2.454(0.028)$ & $\bm{2.452(0.031)}$ & $2.463(0.030)$ \\
		\hline
		$(C):$ $\beta_0 = 1, \beta_{29} = 0.7$ & $4000:1000$ &$3.126(0.018)$ & $3.126(0.018)$ & $\bm{3.108(0.014)}$ \\
		 & $4500:500$ &$3.123(0.024)$ & $3.123(0.024)$ & $\bm{3.106(0.021)}$ \\
		\hline
		$(D):$ $\beta_0 = 1, \beta_1 = -0.5, \beta_7 = 0.5$ & $4000:1000$ &$\bm{1.870(0.016)}$ & $\bm{1.870(0.016)}$ & $1.882(0.024)$ \\
		 & $4500:500$ &$\bm{1.876(0.020)}$ & $\bm{1.876(0.020)}$ & $1.885(0.017)$ \\
		\hline
		$(E):$ $\beta_0 = 1, \beta_{19} = -0.5, \beta_{29} = 0.5$ & $4000:1000$ &$1.873(0.015)$ & $1.873(0.015)$ & $\bm{1.857(0.017)}$ \\
		 & $4500:500$ &$1.869(0.018)$ & $1.869(0.018)$  & $\bm{1.852(0.020)}$\\
		\hline
		$(F):$ $\beta_0 = 1, \beta_1 = -0.5, \beta_7 = -0.5, \beta_{19} = 0.5$ & $4000:1000$ &$1.683(0.013)$ & $1.683(0.013)$ & $\bm{1.631(0.009)}$ \\
		 & $4500:500$ &$1.689(0.017)$ & $1.689(0.017)$  & $\bm{1.635(0.012)}$\\
		\hline
	\end{tabular}}
	\caption{Mean log predictive scores (with standard deviations in brackets) for Bayesian Poisson autoregressive models and our Bayesian tensor factorisations model (BTF) based on 10 Poisson autoregression generated data sets for each one of 6 Scenarios. AIC and BIC columns indicate that the best model has been chosen with the corresponding criterion. Models with the best performance are highlighted in bold.}
	\label{tab:Comp_logLC}
\end{table}

Next, we generated data in which past values affect current random variables in a non-linear fashion as follows. 
There are $\mathcal{K}$ important 
	lag(s) $\{ y_{t-i_{1}}, \dots, y_{t-i_{\mathcal{K}}}  \}$ and, for given  $\nu_+$, $\nu_-$, if $\sum^{\mathcal{K}}_{j = 1} y_{t-i_{j}}  \ge \mathcal{K} \nu_+	$, 
	then $y_{t} \sim \textrm{Poisson}(\nu_+)$; else $y_{t} \sim \textrm{Poisson}(\nu_-)$.	We designed 6 scenarios and the results are shown in Table \ref{tab:Comp_NLC_Large}. 
	Our proposed modelling formulation outperforms the Bayesian Poisson autoregressive model in all but one scenario. 

\begin{table}[h!]\centering
	\ra{1}
	\scalebox{0.88}{
		\begin{tabular}{@{}llllllllllll@{}}\toprule
		& & \multicolumn{2}{c}{Bayesian Poisson autoregression} & \multicolumn{1}{c}{BTF} \\
		\cmidrule{3-4}
		Scenarios & Data Sizes & \multicolumn{1}{c}{AIC} & \multicolumn{1}{c}{BIC} \\ \midrule
		$(A):$ $\nu_+ = 30$, $\nu_- = 50$ & $4000:1000$ &$\bm{3.860(0.032)}$ & $\bm{3.860(0.032)}$ & $3.956(0.251)$ \\
		Important lag: $y_{t-1}$ & $4500:500$ &$\bm{3.869(0.029)}$ & $\bm{3.869(0.029)}$ & $3.982(0.181)$ \\
		\hline
		$(B):$ $\nu_+ = 30$, $\nu_- = 50$ & $4000:1000$ &$3.892(0.056)$ & $3.890(0.055)$ & $\bm{3.691(0.155)}$ \\
		Important lag: $y_{t-7}$ & $4500:500$ &$3.897(0.064)$ & $3.897(0.064)$ & $\bm{3.724(0.173)}$ \\
		\hline
		$(C):$ $\nu_+ = 20$, $\nu_- = 100$ & $4000:1000$ &$3.615(0.207)$ & $3.615(0.207)$ & $\bm{3.437(0.078)}$ \\
		Important lags: $y_{t-3}, y_{t-7}$ & $4500:500$ &$3.665(0.225)$ & $3.668(0.222)$ & $\bm{3.448(0.113)}$ \\
		\hline
		$(D):$ $\nu_+ = 20$, $\nu_- = 100$ & $4000:1000$ &$3.857(0.172)$ & $3.858(0.172)$ & $\bm{3.489(0.088)}$ \\
		Important lags: $y_{t-7} , y_{t-9}$ & $4500:500$ &$3.822(0.192)$ & $3.820(0.187)$ & $\bm{3.470(0.102)}$ \\
		\hline
		$(E):$ $\nu_+ = 20$, $\nu_- = 100$ & $4000:1000$ &$3.426(0.030)$ & $3.426(0.030)$ & $\bm{3.380(0.057)}$ \\
		Important lags: $y_{t-3} , y_{t-7} , y_{t-9}$ & $4500:500$ &$3.440(0.023)$ & $3.441(0.024)$  & $\bm{3.396(0.089)}$\\
		\hline
		$(F):$ $\nu_+ = 20$, $\nu_- = 100$ & $4000:1000$ &$5.338(0.092)$ & $5.338(0.092)$ & $\bm{3.772(0.130)}$ \\
		Important lags: $y_{t-7} , y_{t-8} , y_{t-9}$ & $4500:500$ &$5.270(0.120)$ & $5.270(0.120)$  & $\bm{3.692(0.164)}$\\
		\hline
	\end{tabular}}
	\caption{Mean log predictive scores (with standard deviations in brackets) for Bayesian Poisson autoregressive models and our Bayesian tensor factorisations model (BTF) based on 10 nonlinear generated data sets for each one of 6 Scenarios. AIC and BIC columns indicate that the best model has been chosen with the corresponding criterion. Models with the best performance are highlighted in bold.}
	\label{tab:Comp_NLC_Large}
\end{table}

Finally, we replicated the last exercise by testing the models in a more challenging data generation mechanism in which the response is multivariate. We designed 6 different scenarios by generating an $M$-dimensional time series $\{y_{m,t}\}_{m \in {[1, M]}}$ and assuming that we are interested in predicting   $y_{1, t}$. For $t \leq 10$, we generated $y_{m,t}$ from Pois($\nu_{-}$) for each $m$; for $t > 10$, if $\sum^{\mathcal{K}}_{i = 1} y_{m_i, t-j_i}  \ge \nu_{-}$ we generate $y_{1, t} \sim \textrm{Poisson}(\nu_{+})$, else $y_{1, t} \sim \textrm{Poisson}(\nu_{-})$.	We fitted an $M$-dimensional multivariate Poisson autoregressive model of order $q$ that predicts
$y_{\ell,t}$ with covariates $\{y_{m,t-1}\}_{m \in {M},m\neq \ell}$ as 
	\begin{equation}
	\begin{aligned}
	y_{\ell,t}  &\sim \textrm{Poisson}(\lambda_{\ell,t}),  \\
	\log (\lambda_{\ell,t}) &= \beta_{\ell,0} + \sum_{i= 1}^{q} \beta_{\ell,i} \log (y_{\ell, t-i} + 1) + \sum_{m \neq \ell} 
	\zeta_{\ell,m} y_{m,t-1} 
	\end{aligned}
	\label{multivariate}
	\end{equation}
	where $\beta_{\ell,0},\beta_{\ell,i}$ and $\zeta_{\ell,m}$ are unknown parameters. 
 Table \ref{tab:Comp_NLC_multi} shows that for all 6 Scenarios, the Bayesian tensor factorisation model achieves impressively better predictive performance than  the Bayesian Poisson autoregressive model.

\begin{table}[h!]\centering
	\ra{1}
	\scalebox{0.75}{
		\begin{tabular}{@{}llllllllllll@{}}\toprule
		& & \multicolumn{2}{c}{Bayesian Poisson autoregression} & \multicolumn{1}{c}{BTF} \\
		\cmidrule{3-4}
		Scenarios and non-zero coefficients & Data Sizes & \multicolumn{1}{c}{AIC} & \multicolumn{1}{c}{BIC} \\ \midrule
		$(A):$ $M = 2$; $\nu_{-} = 20$, $\nu_{+} = 10$; & $4000:1000$ &$3.225(0.029)$ & $3.225(0.029)$ & $\bm{3.013(0.037)}$ \\
		Non-zero coefficients for $y_{1, t}$: $y_{1, t-1}$ , $y_{2, t-1}$; & &  \\
		No non-zero coefficient for $y_{2, t}$ & $4500:500$ &$3.243(0.057)$ & $3.243(0.057)$ & $\bm{3.110(0.033)}$ \\
		& &\\
		\hline
		$(B):$ $M = 2$; $\nu_{-} = 20$, $\nu_{+} = 10$; & $4000:1000$ &$2.993(0.030)$ & $2.993(0.030)$ & $\bm{2.705(0.033)}$ \\
		Non-zero coefficients for $y_{1, t}$: $y_{1, t-3}$ , $y_{2, t-5}$; & &  \\
		No non-zero coefficient for $y_{2, t}$ & $4500:500$ &$3.002(0.058)$ & $3.003(0.058)$ & $\bm{2.711(0.040)}$ \\
		& &\\
		\hline
		$(C):$ $M = 2$; $\nu_{-} = 20$, $\nu_{+} = 10$; & $4000:1000$ &$3.488(0.047)$ &$3.487(0.47)$ & $\bm{2.877(0.021)}$ \\
		Non-zero coefficient for $y_{1,t}$: $y_{2, t-1}$; & &  \\
		Non-zero coefficient for $y_{2,t}$: $y_{1, t-2}$ & $4500:500$ &$3.452(0.059)$ &$3.452(0.059)$ & $\bm{2.843(0.024)}$ \\
		& &\\
		\hline
		$(D):$ $M = 2$; $\nu_{-} = 20$, $\nu_{+} = 10$; & $4000:1000$ &$3.207(0.039)$ &$3.206(0.039)$ & $\bm{2.855(0.026)}$ \\
		Non-zero coefficients for $y_{1,t}$: $y_{1, t-3}$ , $y_{2, t-4}$; & &  \\
		Non-zero coefficients for $y_{2,t}$: $y_{1, t-1}$ , $y_{2, t-3}$ , $y_{2, t-5}$ & $4500:500$ &$3.159(0.044)$ &$3.159(0.044)$ & $\bm{2.797(0.029)}$ \\
		& &\\
		\hline
		$(E):$ $M = 3$; $\nu_{-} = 20$, $\nu_{+} = 10$; & $4000:1000$ &$3.632(0.052)$ &$3.632(0.052)$ & $\bm{2.903(0.033)}$ \\
		Non-zero coefficient for $y_{1,t}$: $y_{2, t-1}$; & &  \\
		Non-zero coefficient for $y_{2,t}$: $y_{3, t-2}$; & $4500:500$ &$3.622(0.044)$ &$3.622(0.044)$ & $\bm{2.772(0.030)}$ \\
		Non-zero coefficient for $y_{3,t}$: $y_{1, t-3}$ & & \\
		\hline
		$(F):$ $M = 3$; $\nu_{-} = 60$, $\nu_{+} = 20$; & $4000:1000$ &$6.117(0.149)$ &$6.117(0.149)$ & $\bm{3.508(0.227)}$ \\
		Non-zero coefficients for $y_{1,t}$: $y_{1, t-3} , y_{2, t-4} , y_{3, t-1}$; & &  \\
		Non-zero coefficients for $y_{2,t}$: $y_{1, t-1} , y_{2, t-2} , y_{3, t-5}$; & $4500:500$ &$6.306(0.202)$ &$6.306(0.202)$ & $\bm{3.574(0.173)}$ \\
		Non-zero coefficients for $y_{3,t}$: $ y_{1 ,t-3} , y_{2, t-2} , y_{3, t-5}$ & & \\
		\hline
		
	\end{tabular}}
	\caption{Mean log predictive scores (with standard deviations in brackets) for Bayesian Poisson autoregressive models and our Bayesian tensor factorisations model (BTF) based on 10 nonlinear generated data sets for each one of 6 Scenarios. AIC and BIC columns indicate that the best model has been chosen with the corresponding criterion. Models with the best performance are highlighted in bold.}
	\label{tab:Comp_NLC_multi}
\end{table}	

The times needed to run the MCMC algorithms for Bayesian Poisson autoregressive and BTF models are comparable. For 1000 iterations we needed, on average, 20 seconds for the BTF model implemented with  our matlab code and 25 seconds for the Bayesian Poisson autoregressive models implemented with rjags.  
\section{Applications}
\subsection{Univariate  flu data}
\begin{figure}[h!]
	\centering
	\includegraphics[width=1\linewidth]{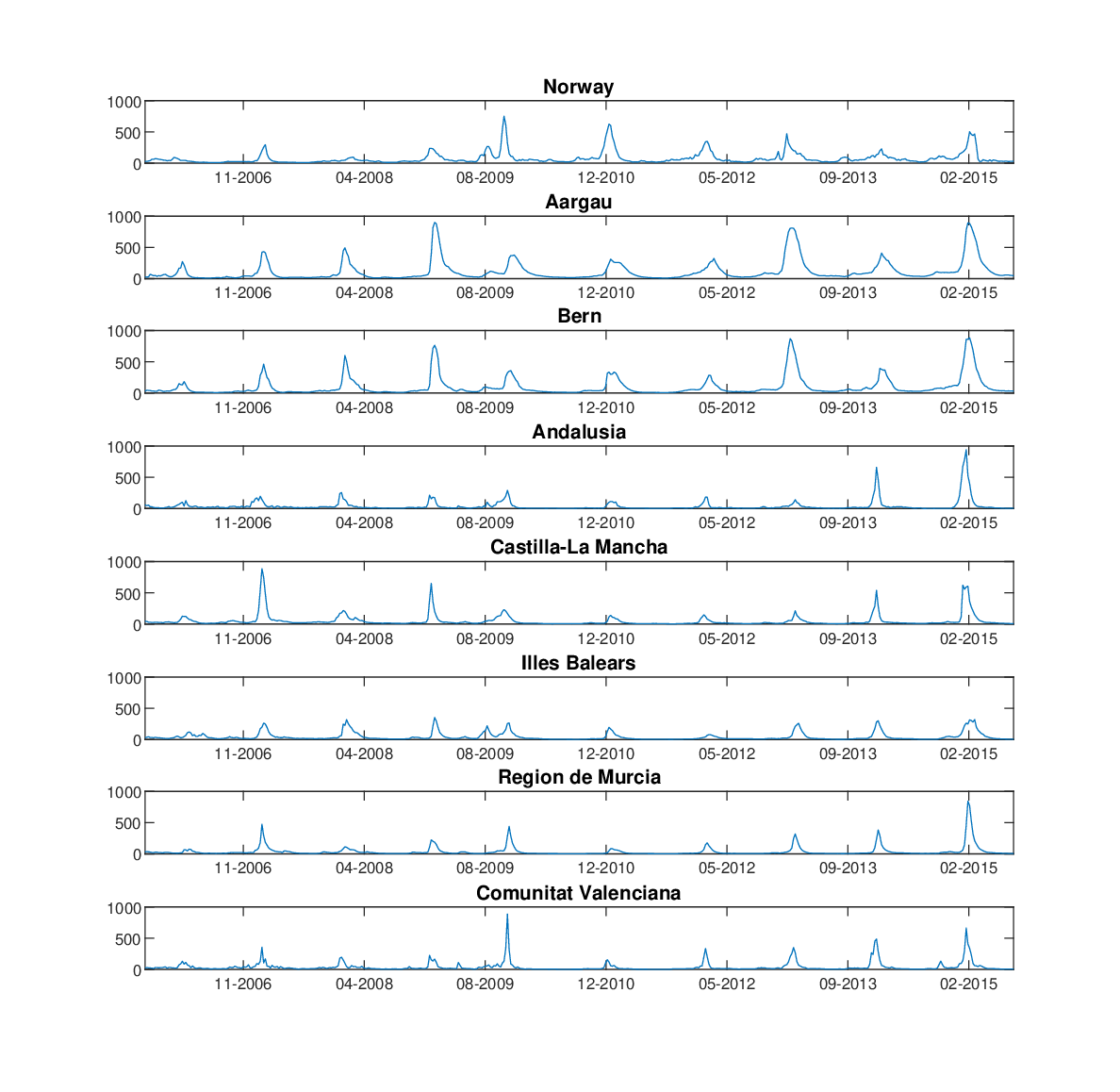}
	\caption{Trace plot of 514 time-series data points counting flu cases in Norway, Aargau as well as Bern in Switzerland and five regions in eastern Spain including Andalusia, Castilla-La Mancha, Illes Balears, Region de Murcia and Valencian Community counted by each week from 09-Oct-2005 to 09-Aug-2015.}
	\label{flu_esp_five}
\end{figure}

We compared our Bayesian tensor factorisation model to Bayesian Poisson autoregressive model with two datasets from Google Flu Trends that refer to 514 Norway, Switzerland  and Castilla–La Mancha weekly flu counts in Spain, see Figure \ref{flu_esp_five}.  We chose the maximum lag $q$ to be 10 for all models we applied to the data.
We examined the sensitivity to the size of the  pre-training data  by considering three scenarios.  We used 103(20\%), 154(30\%) and 206(40\%)  pre-training sizes and compared their predictive ability against the best models for  Bayesian Poisson autoregression formulations  based on AIC and BIC criteria. The last 103 and 52 data points were chosen for out-of-sample test comparison for each dataset.   To demonstrate how our methodology works, we will present MCMC results for the Norway dataset based on 154 training points; results for both datasets and for all training sizes are given at the end of the Section.

%\begin{figure}[h!]
%	\centering
%	\includegraphics[width=1\linewidth]{flu_Nor_Esp.eps}
%	\caption{Trace plot of 514 time-series data points counting flu cases in Norway and five regions in eastern Spain including Andalusia, Castilla-La Mancha, Illes Balears, Region de Murcia and Valencian Community counted by each week from 09-Oct-2005 to 09-Aug-2015.}
%	\label{flu_esp_five}
%\end{figure}

\begin{figure}[h]
	\begin{subfigure}{.45\textwidth}
		\centering
		% include first image
		\includegraphics[width=1\linewidth]{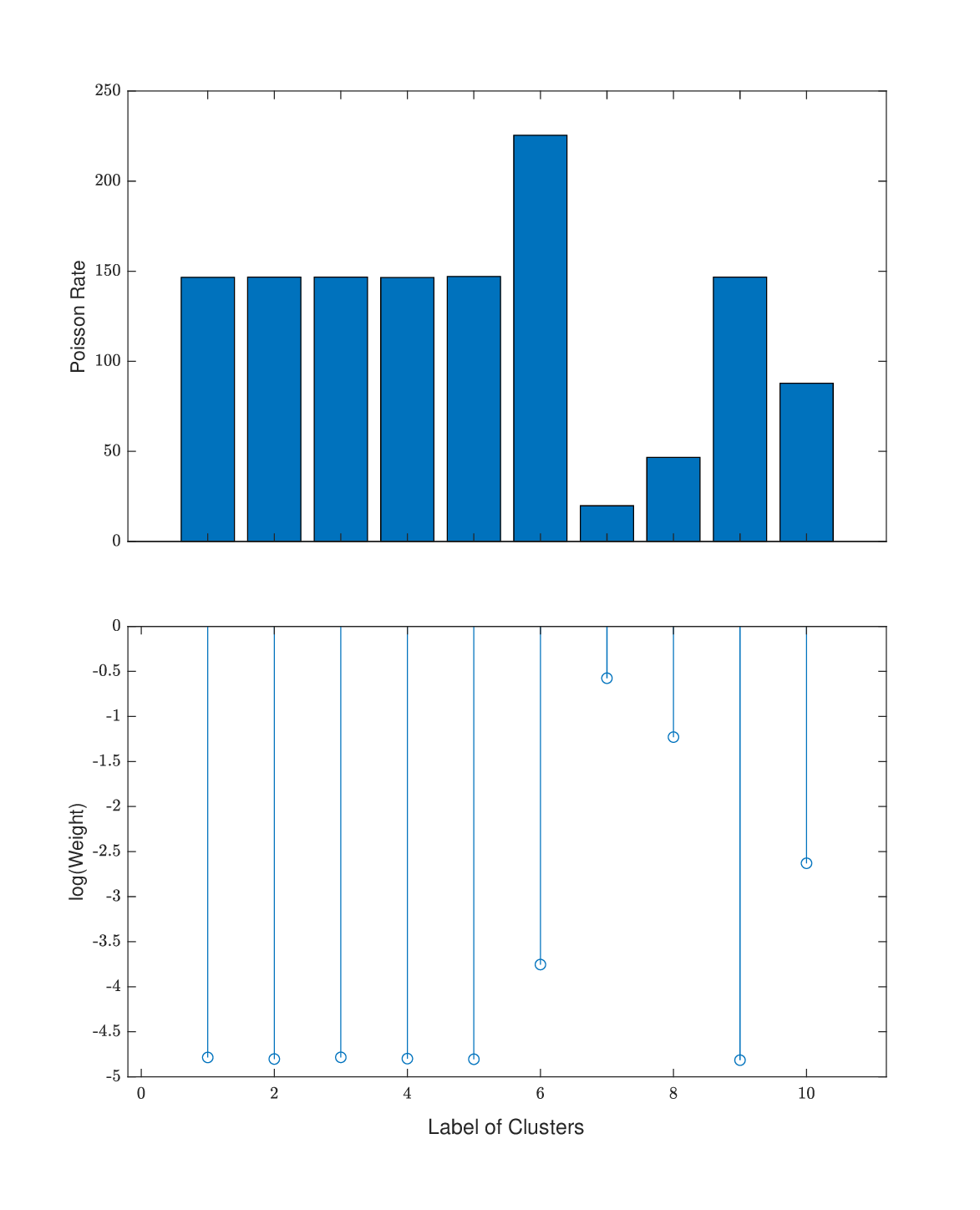}  
		\caption{Total number of clusters to be 10}
		\label{fig:sub-first}
	\end{subfigure}
	\begin{subfigure}{.45\textwidth}
		\centering
		% include second image
		\includegraphics[width=1\linewidth]{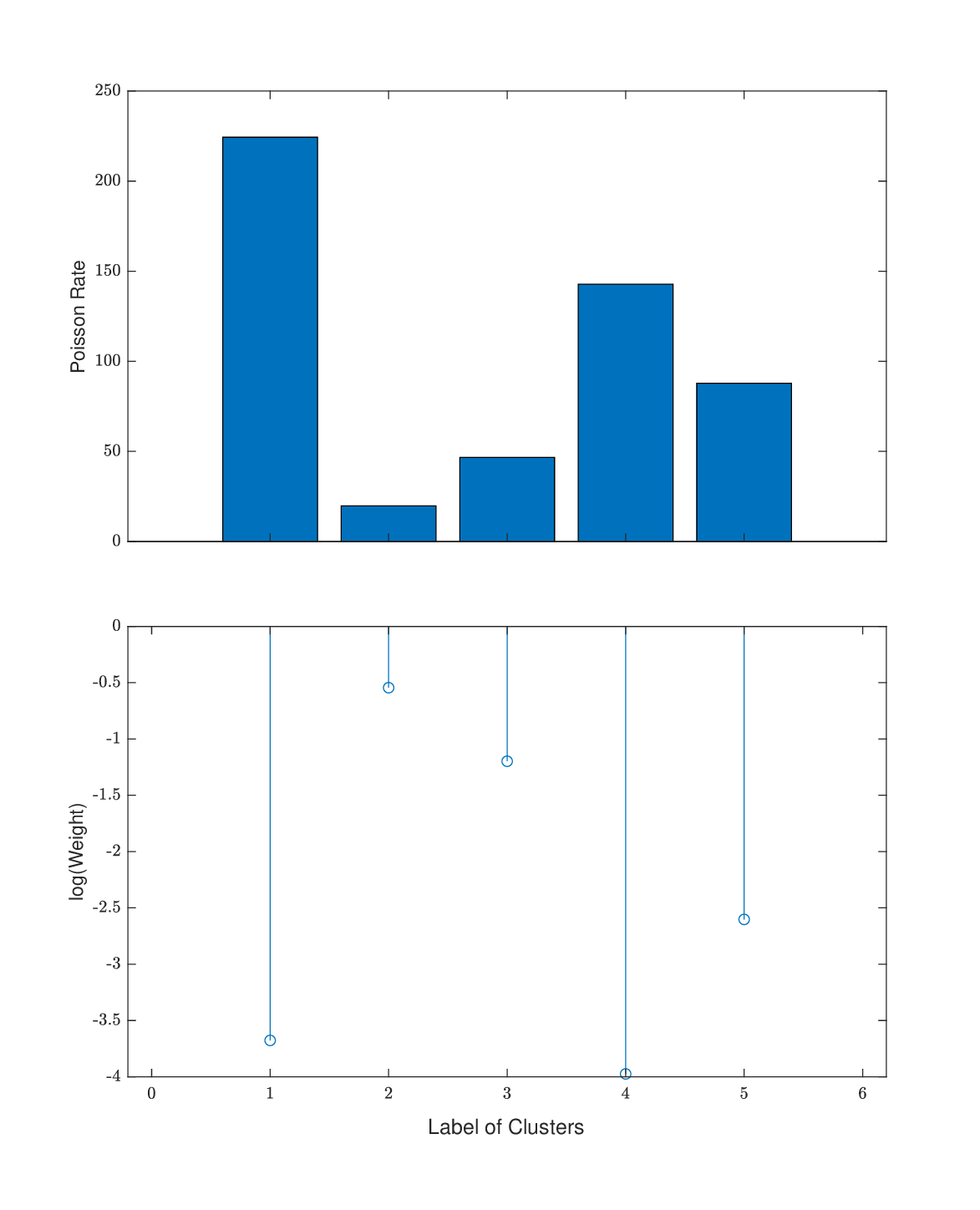}  
		\caption{Total number of clusters to be 5}
		\label{fig:sub-second}
	\end{subfigure}
	\caption{Fitting of  a mixture of Poisson distributions. The dataset used is the pre-training data from flu cases in Norway counted by each week from 09-Oct-2005 to 09-Aug-2015. Panels (a) and (b) indicate that the outcome for a total number of clusters $c$ are 10 and 5 respectively. The top panels illustrate the Poisson rates of their corresponding label of clusters, whilst the bottom panels show their corresponding log weights.}
	\label{MP_Norway}
\end{figure}

\begin{figure}[h]
	\centering
	\includegraphics[width=0.8\linewidth]{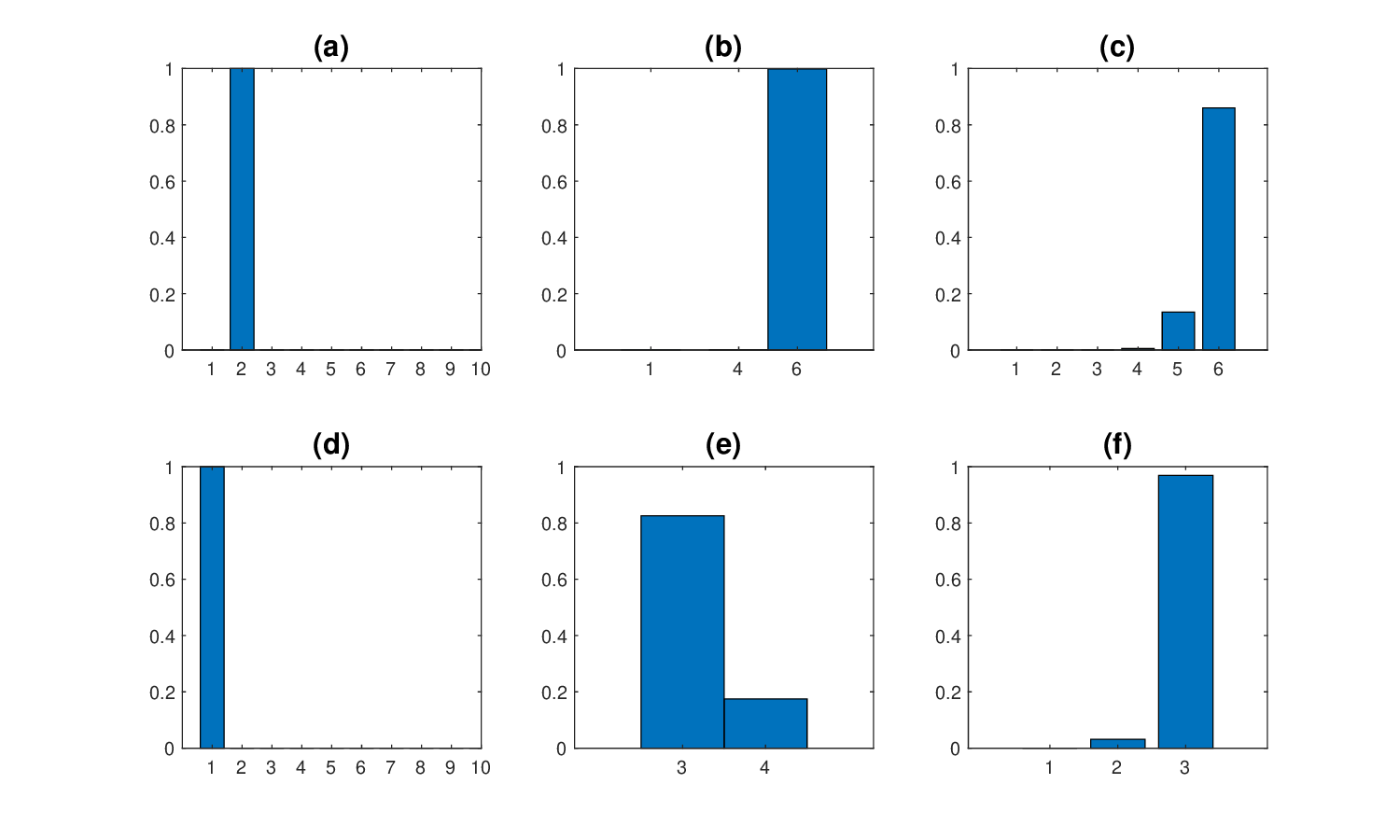}
	\caption{MCMC frequency results. In all panels, the $x$-axis represents the number and the $y$-axis does the relative frequency.Top three panels: 411 training data points;   bottom three panels: 462 training data points. (a,d): The relative frequency distributions for the number of important predictor(s). (b,e): The relative frequency distributions of $\prod_{j=1}^q k_j$, or the total number of possible combinations of $(h_1, \dots, h_q)$. (c,f): The relative frequency distributions of the number of  unique $\lambda_{h_1, \dots, h_q}$. } 
	\label{MCMC_Norway}
\end{figure}

The pre-training results  are illustrated in Figure \ref{MP_Norway}. There are barely significant differences among 6 of the 10 clusters in the left panel so we fix the number of clusters to be 5, see Figure \ref{MP_Norway}. Figure \ref{MCMC_Norway} shows some MCMC results for the rest of our Bayesian tensor factorisation model. With 411 training data points, Panels (a),(b) and (c)  provide strong evidence that there are two important predictors, $6$ possible combinations of $(h_1,...,h_q)$ and $6$ unique $\lambda_{h_1, \dots, h_q}$. Similarly, when the  length of the training dataset is 462 panels (d),(e) and (f)  indicate that there is evidence for only one important predictor, the total number of possible combinations of $(h_1,...,h_q)$ is either 3 or 4, and that there are 3 unique $\lambda_{h_1, \dots, h_q}$.

\begin{figure}[h!]
		\centering
		\includegraphics[width=1\linewidth]{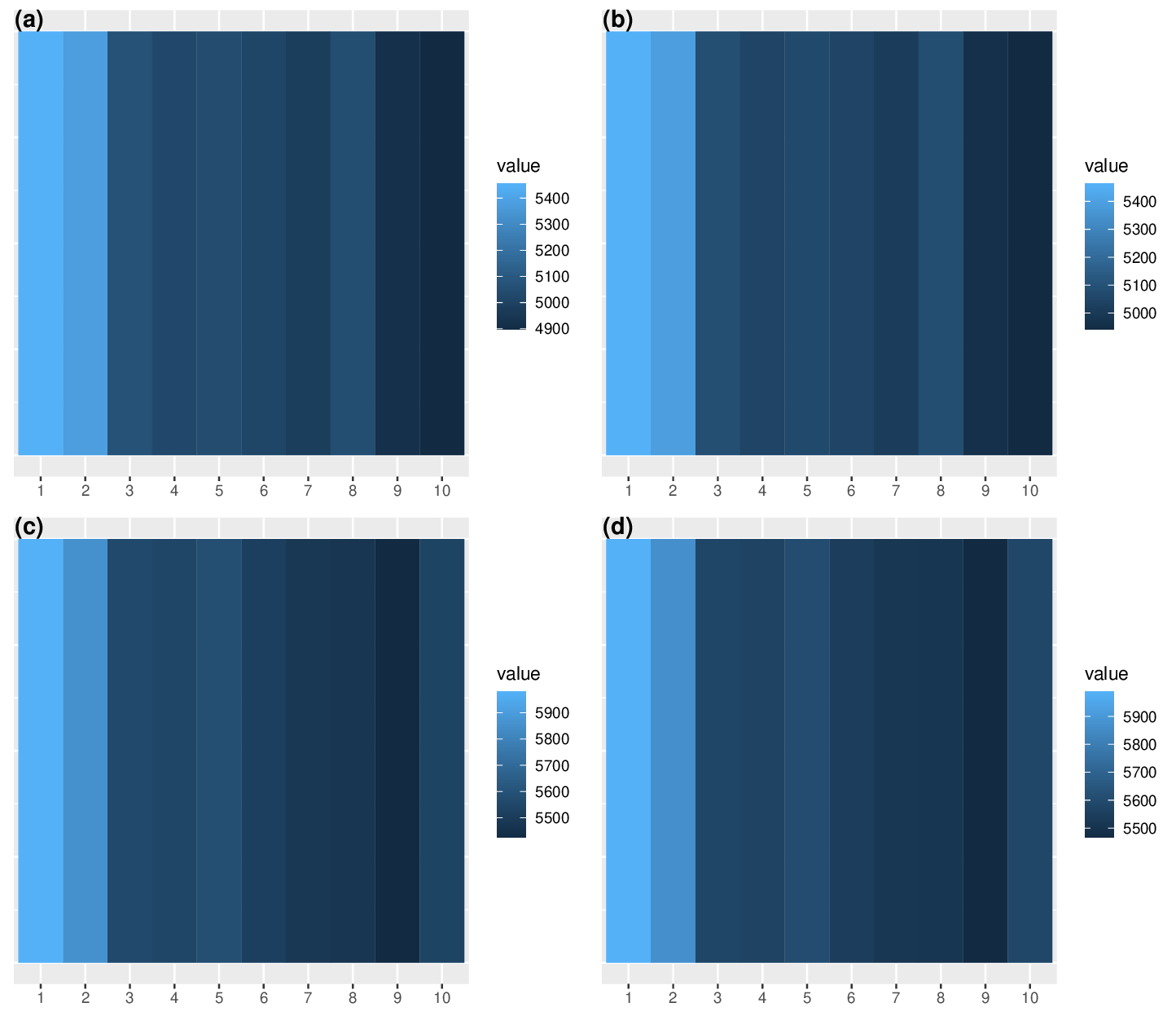}
		\caption{AIC and BIC scores given by PAR($q$) models with q labelled in the x-axis for flu cases in Norway counted by each week from 09-Oct-2005 to 09-Aug-2015. (a): The AIC scores for the scenario with 411 training data points and 103 testing data points; (b): The BIC scores for the scenario with 411 training data points and 103 testing data points; (c): The AIC scores for the scenario with 462 training data points and 52 testing data points; (d): The BIC scores for the scenario with 462 training data points and 52 testing data points.}
		\label{Norway_AB}
	\end{figure}

Model selection results for the Poisson autoregression models are illustrated in Figure \ref{Norway_AB}. MCMC was based on 5,000 burn-in and 10,000 runs by using 'rjags', see \cite{plummer2016rjags}.   For the resulting parameter estimates see Table \ref{tab:PAR_output}.

\begin{table}[h!]\centering
	\ra{1}
	\scalebox{1}{
		\begin{tabular}{@{}clrlrrllllll@{}}\toprule
		 & \textsl{Sizes} & \multicolumn{1}{c}{$411: 103$} & & \multicolumn{1}{c}{$462: 52$} \\ 
		& \textsl{Model selected} & \multicolumn{1}{c}{PAR($10$)} && \multicolumn{1}{c}{PAR($9$)} \\
		\textsl{Coefficient} &      \\ \midrule \midrule
		$\beta_0$ && 0.252(0.035)  && 0.225(0.033)  \\
		$\beta_1$ && 1.472(0.016)  && 1.547(0.023)  \\
		$\beta_2$ && -0.395(0.019) && -0.634(0.038) \\
		$\beta_3$ && -0.172(0.027) && -0.048(0.029) \\
		$\beta_4$ && -0.005(0.022) && 0.071(0.021) \\
		$\beta_5$ && -0.006(0.028) && 0.103(0.021) \\
		$\beta_6$ && 0.050(0.023)  && -0.204(0.017) \\
		$\beta_7$ && -0.111(0.028) && 0.018(0.021) \\
		$\beta_8$ && 0.168(0.029)  && 0.196(0.018) \\
		$\beta_9$&& 0.111(0.029)  && -0.105(0.014) \\
		$\beta_{10}$ && -0.179(0.025) &&  \\
		\bottomrule
	\end{tabular}
}
	\caption{Means of coefficients (with standard deviations in brackets) based on 5,000 burn-in and 10,000 MCMC runs. Two scenarios with different sizes of training  against testing data are shown in each columns with their corresponding selected models indicated.}
	\label{tab:PAR_output}
\end{table}

\clearpage

\begin{figure}[h!]
	\centering
	\includegraphics[width=1\linewidth]{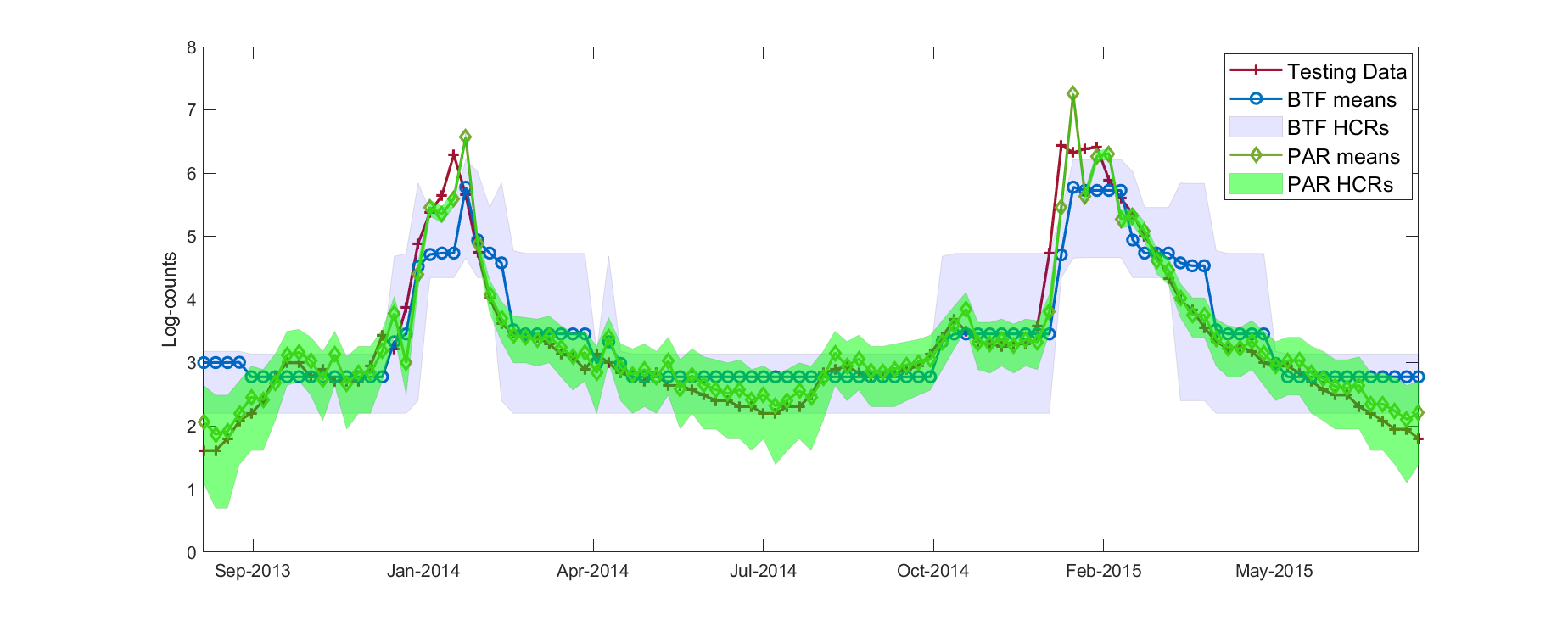}
	\caption{Out of sample predictive means and $95\%$ highest credible regions (HCRs)  of Bayesian tensor factorisations (BTF) and  Poisson autoregressive models (PAR) compared against the Castilla–La Mancha data. The sizes of training and testing data are 411 and 103 respectively.}
	\label{80_CLM_log}
\end{figure}

Table \ref{tab:Comp_Norway_CLM} indicates that in all pre-training size scenarios BTF  outperform, in terms of predictive ability expressed with log predictive scores, Bayesian Poisson autoregression models.  There is clearly a trade-off between good mixture estimation and adequate training size that is expressed in  small and large pre-training sizes respectively.  In our small empirical study it seems that there is evidence for some robustness in the inference procedure when the pre-training size is small, since 103 points outperform 206 points with the 154 points being the best performing pre-training size. The predictive means and $95\%$ credible intervals of BTF and of the PAR(5) model that had one of the best predictive performances based on  103 test data are depicted in Figure \ref{80_CLM_log}. 

\begin{table}[h!]\centering
	\ra{1}
	\scalebox{0.78}{
		%\begin{tabular}{@{}llllllllllll@{}}\toprule
		\begin{tabular}{@{}cccccccccccc@{}}\toprule
		& & \multicolumn{2}{c}{Poisson autoregression} & \multicolumn{3}{c}{BTF} \\
		\cmidrule{3-4} \cmidrule{5-7}
		Country/Region & Data Sizes & AIC & BIC & 103 PTDPs& 154 PTDPss & 206 PTDPs  \\ \midrule
		Norway & $411:103$ & $7.560(10)$ & $7.560(10)$ & $6.054$ & $\bm{5.846}$ & $6.440$\\
		 & $462:52$ &$7.805(9)$ & $7.805(9)$ & $6.110$ & $\bm{6.079}$ & $6.289$\\
%		\hline
%            Bern, Switzerland  & $411:103$ & $6.584(10)$ & $6.584(10)$ & $5.460$ & $\bm{5.312}$ & $7.029$\\
%		 & $462:52$ &$7.476(10)$ & $7.476(10)$ & $6.222$ & $\bm{6.059}$ & $6.980$\\
		\hline
		Castilla-La Mancha, Spain & $411:103$ & $12.295(10)$ & $12.607(5)$ & 5.667 & $\bm{5.416}$ & 6.760 \\
		 & $462:52$ &$15.073(9)$ & $15.073(9)$ & 5.858 & $\bm{5.664}$ & 6.268 \\
		\hline
	\end{tabular}}
	\caption{Log predictive scores for Bayesian Poisson autoregression models and our Bayesian tensor factorisations model (BTF) for flu counts datasets in Norway and Castilla-La Mancha, Spain.	  The BTF model has performed with 103, 154 and 206 pre-training data points (PTDPs).  AIC and BIC columns indicate that the best model has been chosen (in brackets) with the corresponding criterion.  Training and testing data sizes appear in the second column. Models with the best performance are highlighted in bold.}
	\label{tab:Comp_Norway_CLM}
\end{table}

The average run times for the MCMC algorithms  for BTF and the Bayesian Poisson autoregression models are comparable.  For the former, 1000 iterations take approximately 20 seconds with our code written in matlab, whereas the latter  takes approximately 25 seconds for 1000 iterations in the R package ‘rjags’.

\subsection{Multivariate flu data}
We revisit the flu data of the previous subsection by jointly modelling flu cases in  (i) the adjacent Swiss cantons of Bern and  Aargau and (ii) in  five neighbouring regions in south-eastern Spain, namely Andalusia, Castilla-La Mancha, Illes Balears, Region de Murcia and Valencian Community. 
The data are depicted in Figure \ref{flu_esp_five} and consist of 514 weekly counts from  09-Oct-2005 to 09-Aug-2015.

We chose the maximum lag $q$ to be ten for all multivariate BTF models  we applied to the data. 
The sizes of training against the testing dataset are $411:103$ and $462:52$ respectively. Our BTF considered the first 154 data points as the pre-training dataset.

\begin{figure}[h!]
	\centering
	\includegraphics[width=0.45\linewidth]{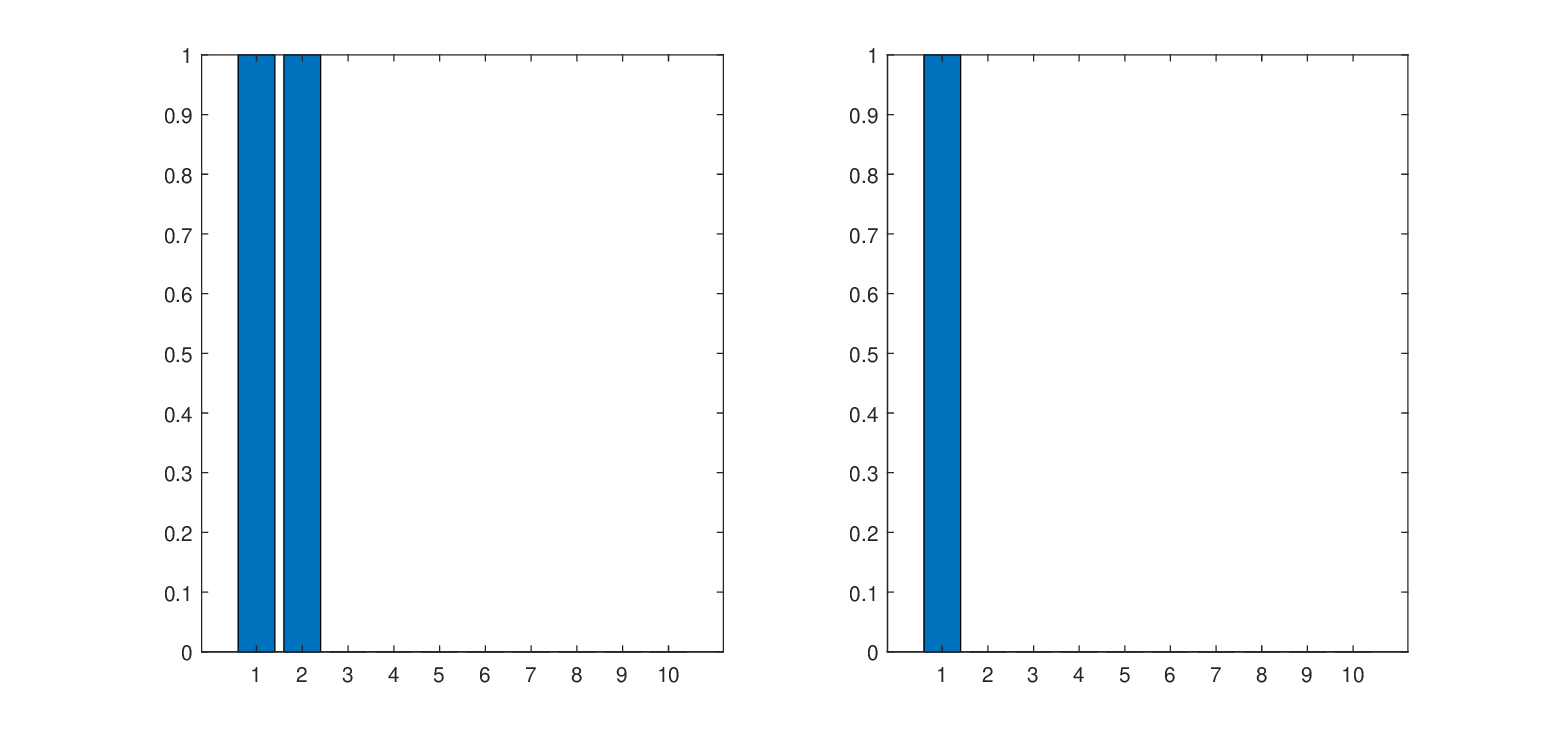}
	\includegraphics[width=0.45\linewidth]{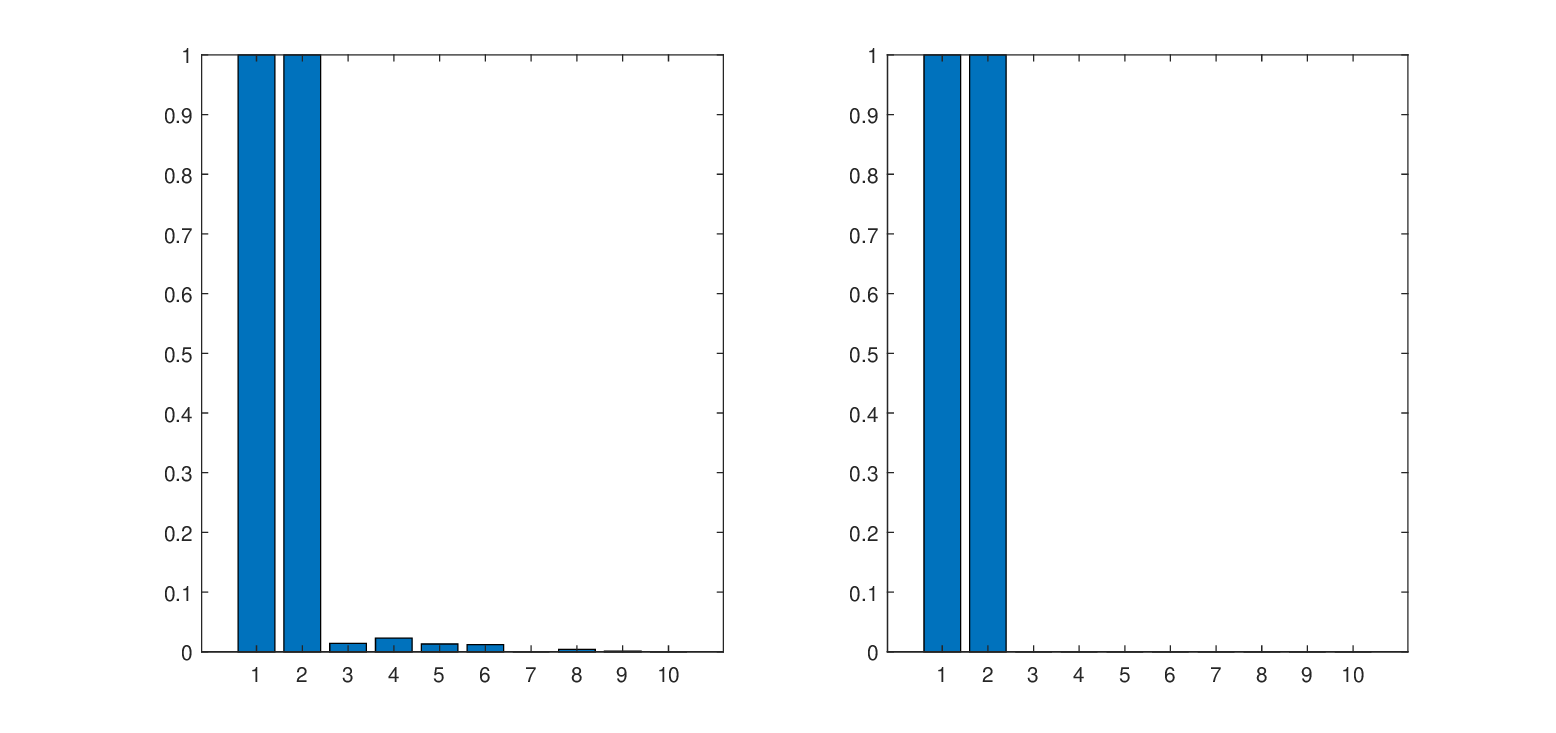}
	\caption{Lag selection for the Norway (left pair) and the Castilla-La Mancha (right pair) flu datasets. Each pair of figures represents (i) the inclusion proportions (y-axis) of different lags (x-axis) for the scenario with 411 training  and 103 testing data points and (ii)  the inclusion proportions (y-axis) of different lags (x-axis) for the scenario with 462 training data points and 52 testing data points. }
	\label{flu_RD_CLM}
\end{figure}

\begin{figure}[h]
	\begin{subfigure}{.45\textwidth}
		\centering
		% include first image
		\includegraphics[width=1\linewidth]{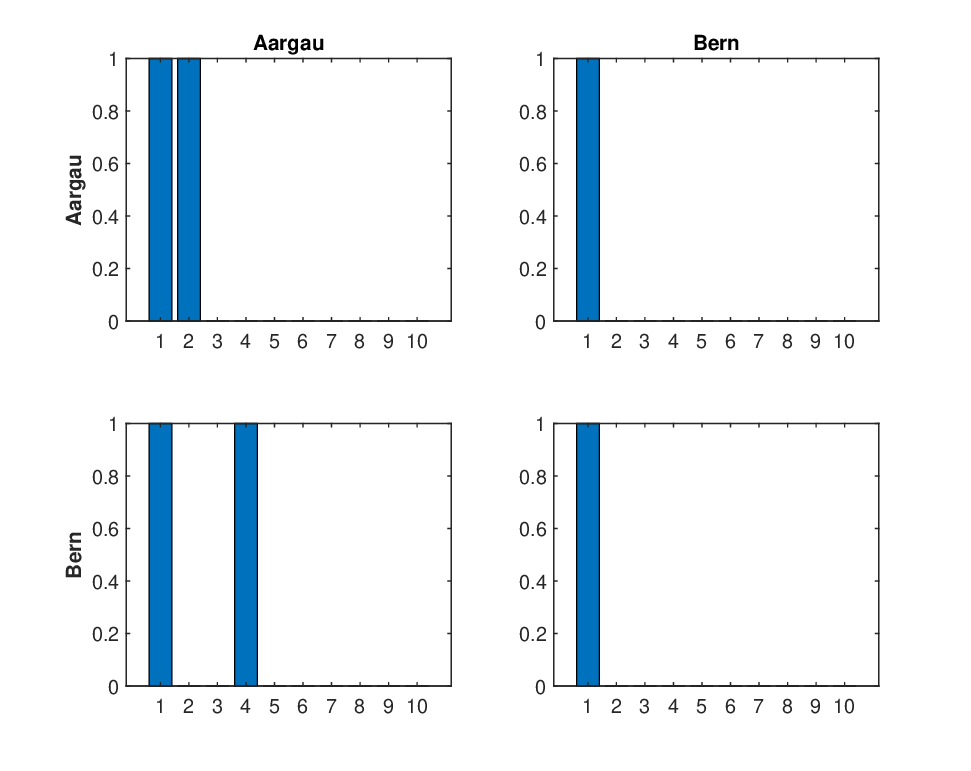}  
		\caption{411 training and 103 testing points}
		\label{fig:sub-first}
	\end{subfigure}
	\begin{subfigure}{.45\textwidth}
		\centering
		% include second image
		\includegraphics[width=1\linewidth]{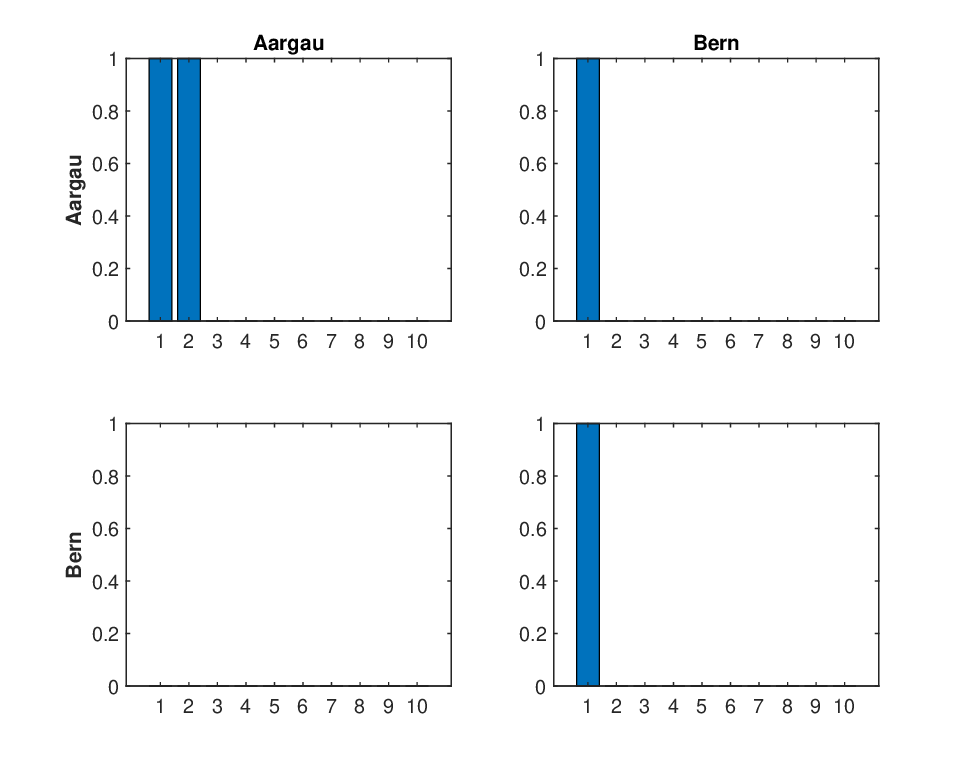}  
		\caption{462 training and 52 testing points }
		\label{fig:sub-second}
	\end{subfigure}
	\caption{Important lag selection for the Swiss flu dataset. Y-axis represents the inclusion proportions of different lags in x-axis.}
	\label{MP_Swiss}
\end{figure}

\begin{figure}[h!]
	\centering
	\includegraphics[width=1\linewidth]{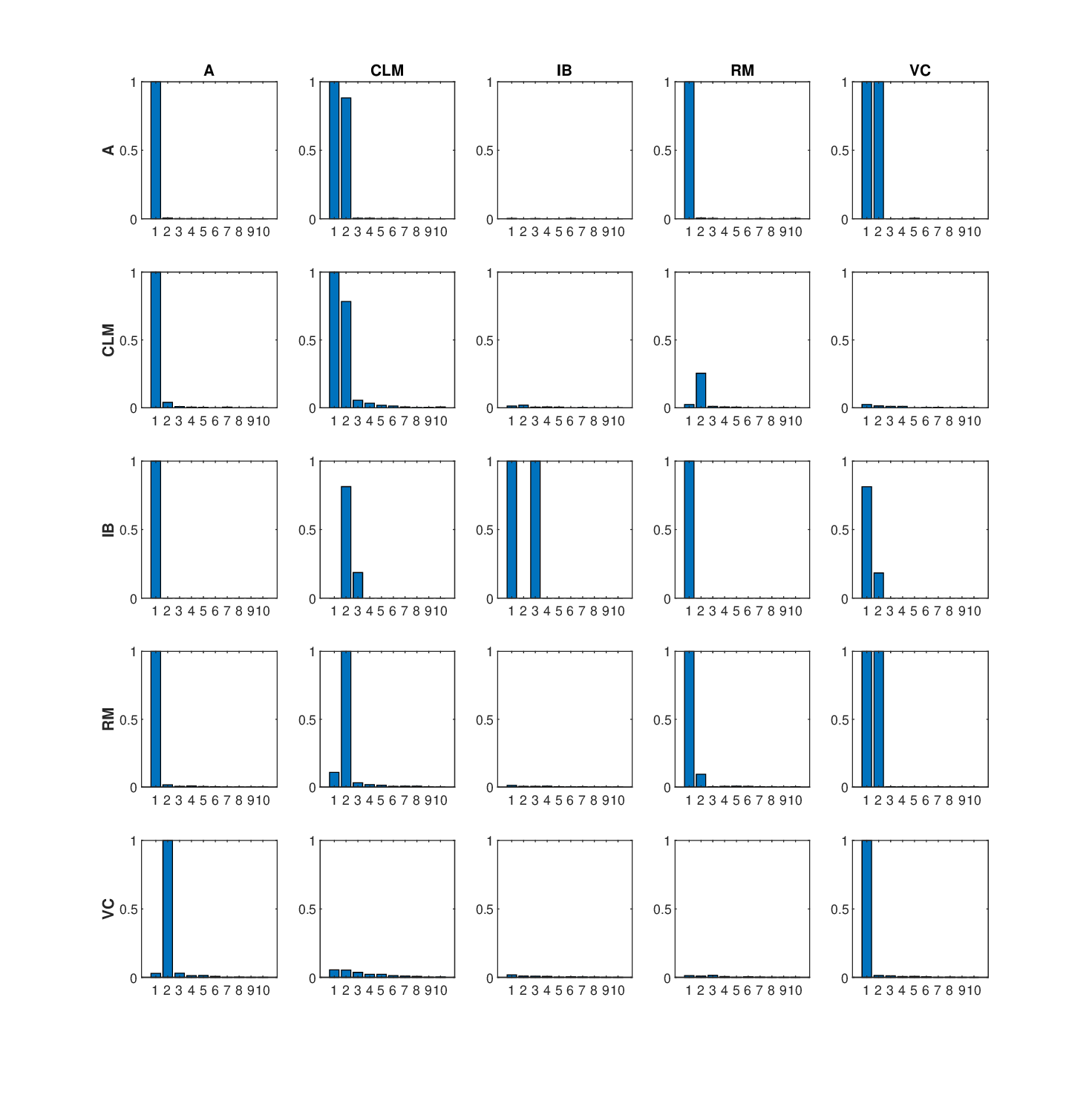}
	\caption{Important lag selection for the south-eastern Spain flu dataset. Y-axis represents the inclusion proportions of different lags in x-axis for the scenario with 411 training data points and 103 testing data points.  A: Andalusia; CLM: Castilla-La Mancha; IB: Illes Balears; RM: Region de Murcia; VC: Valencian Community.}
	\label{flu_RD_esp_82}
\end{figure}

\begin{figure}[h!]
	\centering
	\includegraphics[width=1\linewidth]{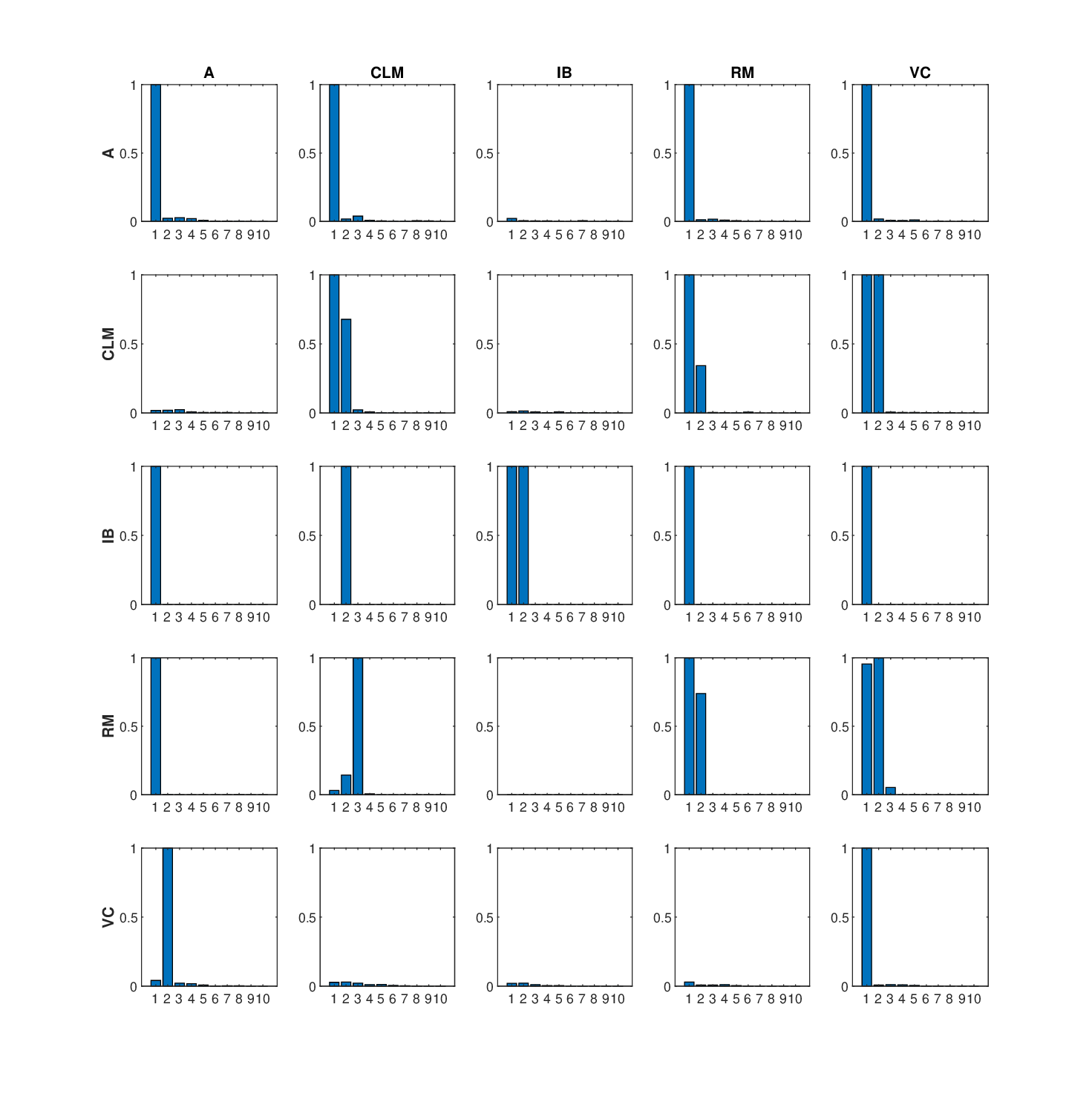}
	\caption{Important lag selection for the south-eastern Spain flu dataset. Y-axis represents the inclusion proportions of different lags in x-axis for the scenario with 462 training data points and 52 testing data points.  A: Andalusia; CLM: Castilla-La Mancha; IB: Illes Balears; RM: Region de Murcia; VC: Valencian Community.}
	\label{flu_RD_esp_91}
\end{figure}

Figures \ref{flu_RD_CLM}-\ref{flu_RD_esp_91} illustrate how lags were selected  in each real data application. Note that a lag is considered to be important, and thus is selected,  when its corresponding relative frequency distribution is higher than 0.5. 

The predictive ability of the models compared to the Bayesian Poisson autoregression models are given  in Tables \ref{tab:Comp_flu_swiss} and \ref{tab:Comp_flu_esp}.   In the Swiss cantons it seems that the BTF model underperforms when Aargau flu cases are predicted from past flu cases of Aargu and Bern, whereas it outperforms when we predict Bern cases based on past data from Aargau and Bern.   An informal justification of this behaviour is that from the data it seems that the two series have very high positive contemporaneous and lag-one correlations so naturally the model (\ref{multivariate}) that captures very well these linear dependencies outperforms our model.  Such situations are expected when a general non-parametric model is compared to a linear model with the corresponding data generating mechanism to be primarily linear-based.

Table \ref{tab:Comp_flu_esp} presents the five-dimensional example of Spanish regions in which the counts of each region are predicted from past counts of all five regions.  Here, in eight out of ten cases BTF outperforms the Poisson autoregression model and in particular the log-predictive scores are dramatically lower in all cases with smaller training (462) and  higher test (103) sizes. This is not surprising since our model is capable of capturing the complicated five-dimensional dependencies created in these Spanish regions.

\begin{table}[h!]\centering
	\ra{1}
	\scalebox{0.9}{
		\begin{tabular}{@{}cccccccccccc@{}}\toprule
		& & \multicolumn{2}{c}{Bayesian Poisson autoregression} & \multicolumn{1}{c}{BTF} \\
		\cmidrule{3-4}
		Region & Data Sizes & AIC & BIC &  \\ \midrule
		Aargau & $411:103$ & $\bm{4.574(10)}$ & $\bm{4.574(10)}$ & $5.719$ & \\
		 & $462:52$ &$\bm{5.001(10)}$ & $5.041(4)$ & $5.836$ \\
		\hline
		Bern & $411:103$ & $6.155(10)$ & $6.155(10)$ & $\bm{5.328}$ \\
		 & $462:52$ &$6.632(10)$ & $6.632(10)$ & $\bm{6.103}$ \\
		\hline
	\end{tabular}}
	\caption{Log predictive score between Bayesian Poisson autoregressive model and Bayesian tensor factorisations model (BTF) for multivariate flu counts datasets. Multiple datasets include flu counts in two cantons in Switzerland, Aargau and Bern. AIC and BIC columns indicate that the best model has been chosen (in brackets) with the corresponding criterion. Models with the best performance are highlighted in bold.}
	\label{tab:Comp_flu_swiss}
\end{table}

\begin{table}[h!]\centering
	\ra{1}
	\scalebox{0.9}{
		\begin{tabular}{@{}llllllllllll@{}}\toprule
		& & \multicolumn{2}{c}{Bayesian Poisson autoregression} & \multicolumn{1}{c}{BTF} \\
		\cmidrule{3-4}
		Region & Data Sizes & AIC & BIC \\ \midrule
		Andalucia & $411:103$ & $\bm{9.097(10)}$ & $9.296(4)$ & $17.205$ & \\
		 & $462:52$ &$17.550(4)$ & $17.550(4)$ & $\bm{14.006}$ \\
		\hline
		Castilla-La Mancha & $411:103$ & $14.589(7)$ & $14.467(5)$ & $\bm{5.760}$ \\
		 & $462:52$ &$23.765(10)$ & $23.708(9)$ & $\bm{5.992}$ \\
		\hline
		Illes Balears & $411:103$ & $14.334(10)$ & $14.297(3)$ & $\bm{4.335}$ \\
		 & $462:52$ &$6.788(10)$ & $7.019(5)$ & $\bm{5.178}$ \\
		\hline
		Region de Murcia & $411:103$ & $25.593(10)$ & $25.379(8)$ & $\bm{10.665}$ \\
		 & $462:52$ &$\bm{5.771(10)}$ & $\bm{5.771(10)}$ & $15.078$ \\
		\hline
		Valencian Community & $411:103$ & $13.760(10)$ & $14.601(8)$ & $\bm{5.997}$ \\
		 & $462:52$ &$21.532(10)$ & $21.532(10)$ & $\bm{6.450}$ \\
		\hline
	\end{tabular}}
	\caption{Log predictive score between Bayesian Poisson autoregressive model and Bayesian tensor factorisations model (BTF) for multivariate flu counts datasets. Multiple datasets include flu counts in Andalusia, Castilla-La Mancha, Illes Balears, Region de Murcia and Valencian Community.	AIC and BIC columns indicate that the best model has been chosen (in brackets) with the corresponding criterion. Models with the best performance are highlighted in bold.}
	\label{tab:Comp_flu_esp}
\end{table}

\section{Discussion}

We have introduced a new flexible  modelling framework for
that extends Bayesian tensor factorisations to multivariate time series of count data.
Extensive simulation studies and analysis of real data provide evidence that the flexibility of these models offers an important alternative to other multivariate time series models for counts.

An important aspect of our proposed models is that direct MCMC inference cannot avoid an increased computational complexity  as observed counts grow.  We have dealt with this issue with a two-stage inferential procedure that successfully deals with large observed counts.

\section*{Acknowledgements}

We would like to thank Abhar Sarkar for kindly providing us the code for \cite{sarkar2016bayesian}.

\section{Declarations}

\begin{itemize}
\item Funding: Not applicable
\item Conflicts of interest/Competing interests: Not applicable
\item Ethics approval: Not applicable
\item Consent to participate: Not applicable
\item Consent for publication: Not applicable
\item Availability of data and material: The google flu data are publicly available.
\item Code availability: The code will be free and available from Petros Dellaportas' web site.
\item Author contributions: The code has been written by Zhongzhen Wang. Zhongzhen Wang, Petros Dellaportas and Ioannis Kosmidis had equal contributions at the development of the theory.
\item Licence: For the purpose of open access, the authors have
  applied a Creative Commons Attribution (CC BY) licence to any Author
  Accepted Manuscript version arising from this submission.

\end{itemize}
\bibliographystyle{chicago}
\bibliography{btfcounts}

\end{document}